\documentclass{aa}  

\usepackage{graphicx}
\usepackage{graphics,psfrag}
\usepackage{graphicx,psfrag}
\usepackage{txfonts}
\usepackage{tipa}
\usepackage{amsmath}
\usepackage{mathrsfs}
\usepackage{natbib,twoopt}
\usepackage[breaklinks=true]{hyperref} 
\bibpunct{(}{)}{;}{a}{}{,}             
\makeatletter
  \newcommandtwoopt{\citeads}[3][][]{\href{http://adsabs.harvard.edu/abs/#3}%
    {\def\hyper@linkstart##1##2{}%
     \let\hyper@linkend\@empty\citealp[#1][#2]{#3}}}
  \newcommandtwoopt{\citepads}[3][][]{\href{http://adsabs.harvard.edu/abs/#3}%
    {\def\hyper@linkstart##1##2{}%
     \let\hyper@linkend\@empty\citep[#1][#2]{#3}}}
  \newcommandtwoopt{\citetads}[3][][]{\href{http://adsabs.harvard.edu/abs/#3}%
    {\def\hyper@linkstart##1##2{}%
     \let\hyper@linkend\@empty\citet[#1][#2]{#3}}}
  \newcommandtwoopt{\citeyearads}[3][][]%
    {\href{http://adsabs.harvard.edu/abs/#3}
    {\def\hyper@linkstart##1##2{}%
     \let\hyper@linkend\@empty\citeyear[#1][#2]{#3}}}
\makeatother
\begin{document}

   \title{Theoretical modelling of the AGN iron line vs continuum time-lags in the lamp-post geometry}

   \author{A. Epitropakis
          \inst{1}
          \and
          I. E. Papadakis\inst{1,2}
          \and
          M. Dov\v{c}iak\inst{3}
          \and
          T. Pech\'a\v{c}ek\inst{1,3}
          \and
          D. Emmanoulopoulos\inst{4,1}
          \and
          V. Karas\inst{3}
          \and
          I. M. M$^{\mathrm{c}}$Hardy\inst{4}
          }

   \institute{Department of Physics and Institute of Theoretical and Computational Physics, University of Crete, 71003 Heraklion, Greece
         \and
         IESL, Foundation for Research and Technology-Hellas, GR-71110 Heraklion, Crete, Greece
         \and 
         Astronomical Institute, Academy of Sciences, Bo\v{c}n\'{\i}~II 1401, CZ-14131~Prague, Czech Republic
         \and
         Physics and Astronomy, University of Southampton, Southampton, SO17 1BJ, UK
             }
\authorrunning{A. Epitropakis et al.}
\titlerunning{Theoretical modelling of the AGN iron line vs continuum time-lags}
\date{Received .. ...... 2015; accepted .. ...... 2015}


  \abstract
   {Theoretical modelling of time-lags between variations in the Fe K$\alpha$ emission and the X-ray continuum might shed light on the physics and geometry of the X-ray emitting region in active galaxies (AGN) and X-ray binaries. We here present the results from a systematic analysis of time-lags between variations in two energy bands ($5-7$ vs $2-4\,\mathrm{keV}$) for seven X-ray bright and variable AGN.}
   {We estimate time-lags as accurately as possible and fit them with theoretical models in the context of the lamp-post geometry. We also constrain the geometry of the X-ray emitting region in AGN.}
   {We used all available archival \textit{XMM-Newton} data for the sources in our sample and extracted light curves in the $5-7$ and $2-4\,\mathrm{keV}$ energy bands. We used these light curves and applied a thoroughly tested (through extensive numerical simulations) recipe to estimate time-lags that have minimal bias, approximately follow a Gaussian distribution, and have known errors. Using traditional $\chi^2$ minimisation techniques, we then fitted the observed time-lags with two different models: a phenomenological model where the time-lags have a power-law dependence on frequency, and a physical model, using the reverberation time-lags expected in the lamp-post geometry. The latter were computed assuming a point-like primary X-ray source above a black hole surrounded by a neutral and prograde accretion disc with solar iron abundance. We took all relativistic effects into account for various X-ray source heights, inclination angles, and black hole spin values.}
   {Given the available data, time-lags between the two energy bands can only be reliably measured at frequencies between $\sim5\times10^{-5}\,\mathrm{Hz}$ and $\sim10^{-3}\,\mathrm{Hz}$. The power-law and reverberation time-lag models can both fit the data well in terms of formal statistical characteristics. When fitting the observed time-lags to the lamp-post reverberation scenario, we can only constrain the height of the X-ray source. The data require, or are consistent with, a small ($\lesssim10$ gravitational radii) X-ray source height.}
{In principle, the $5-7\,\mathrm{keV}$ band, which contains most of the Fe K$\alpha$ line emission, could be an ideal band for studying reverberation effects, as it is expected to be dominated by the X-ray reflection component. We here carried out the best possible analysis with \textit{XMM-Newton} data. Time-lags can be reliably estimated over a relatively narrow frequency range, and their errors are rather large. Nevertheless, our results are consistent with the hypothesis of X-ray reflection from the inner accretion disc.}
   \keywords{Accretion, accretion disks -  Relativistic processes - Galaxies: active - Galaxies: Seyfert  - X-rays: galaxies  }

   \maketitle
%

\section{Introduction} \label{sec1}

According to the currently accepted paradigm, active galactic nuclei (AGN) contain a central, super-massive ($M_{\mathrm{BH}}\sim10^{6-9}\,\mathrm{M}_{\odot}$) black hole (BH), onto which matter accretes in a disc-like configuration. In the standard $\alpha$-disc model \citep{1973A&A....24..337S}, this accretion disc is optically thick and releases part of its gravitational energy in the form of black-body radiation, which peaks at optical to ultra-violet wavelengths. A fraction of these low-energy thermal photons is assumed to be Compton up-scattered by a population of high-energy ($\sim100\,\mathrm{keV}$) electrons, which is often referred to as the corona. The Compton up-scattered disc photons form a power-law spectrum that is observed in the X-ray spectra of AGN at energies $\sim2-10\,\mathrm{keV}$ \citep[e.g.][]{1991ApJ...380L..51H}. We here refer to this source as the X-ray source and to its spectrum as continuum emission. Depending on the X-ray source and disc geometry, a significant amount of continuum emission may illuminate disc and be reflected towards a distant observer.

The strongest observable features of such a reflection spectrum from neutral material are the fluorescent Fe K$\alpha$ emission line at $\sim6.4\,\mathrm{keV}$ and the so-called Compton hump, which is an excess of emission at energies $\sim10-30\,\mathrm{keV}$ \citep[e.g.][]{1991MNRAS.249..352G}. Additionally, if the disc is mildly ionised, an excess of emission at energies $\sim0.3-1\,\mathrm{keV}$ can be observed \citep[e.g.][]{2005MNRAS.358..211R}. In addition to these spectral features, the X-ray reflection scenario also predicts unique timing signatures. For example, \citet{2016A&A...588A..13P} showed that X-ray reflection should leave its imprint in the X-ray power spectra. Owing to X-ray illumination, the observed power spectra should show a prominent dip at high frequencies, and an oscillatory behaviour, with decreasing amplitude, at higher frequencies. These reverberation echo features should be more prominent in energy bands where the reflection component is more pronounced. Furthermore, as a result of the different light travel paths between photons arriving directly at a distant observer and those reflected off the surface of the disc, variations in the reprocessed disc emission are expected to be delayed with respect to continuum variations. The magnitude of these delays will depend on the size and location (with respect to the disc) of the X-ray source, the viewing angle, the mass, and spin of the BH.

Hints for such reverberation delays were first reported by \citet{2007MNRAS.382..985M} in Ark 564. The first statistically robust detection was later reported by \citet{2009Natur.459..540F} in 1H 0707--495, where variations in the $0.3-1\,\mathrm{keV}$ band (henceforth, the soft band) were found to lag behind variations in the $1-4\,\mathrm{keV}$ band by $\sim30\,\mathrm{sec}$ on timescales shorter than $\sim30\,\mathrm{min}$. The discovery of these time-lags, commonly referred to as soft lags in the literature, has triggered a significant amount of research over the past few years. Soft lags have been discovered in $\sim20$ AGN so far \citep[see e.g.][for a review]{2014A&ARv..22...72U}.

A growing number of AGN show evidence of reverberation time-lags between the Fe K$\alpha$ emission line and the continuum \citep[e.g.][]{2012MNRAS.422..129Z,2013MNRAS.428.2795K,2013MNRAS.430.1408K,2013MNRAS.434.1129K,2013ApJ...767..121Z,2014MNRAS.440.2347M}, and between the Compton hump and the continuum \citep[e.g.][]{2014ApJ...789...56Z,2015MNRAS.446..737K}. Detecting them is a particularly difficult task because of the low sensitivity of most current detectors and the intrinsically low brightness of AGN at Fe K$\alpha$ line and Compton hump energies.

Theoretical modelling of X-ray time-lags can elucidate the physical and geometrical nature of the X-ray emitting region in AGN. This requires knowledge of how the disc responds to the continuum emission, and the construction of theoretical time-lag spectra, which can then be fitted to the observed ones. Initial modelling attempts were based on the assumption that this response is a simple top-hat function \citep[e.g.][]{2011MNRAS.412...59Z,2011MNRAS.416L..94E}. \citet{2012MNRAS.420.1145C} were the first to consider a more realistic scenario, in which relativistic effects and a moving X-ray source were considered to quantify the response of the disc. They deduced that, for 1H 0707--495, a more complex physical model is required to explain both the source geometry and intrinsic variability. More recently, \citet{2013MNRAS.430..247W} considered a variety of different geometries for the primary X-ray source and deduced that, in 1H 0707--495, it has a radial extent of $\sim35r_{\mathrm{g}}$ (where $r_{\mathrm{g}}\equiv GM_{\mathrm{BH}}/c^2$ is the gravitational radius) and is located at a height of $\sim2r_{\mathrm{g}}$ above the disc plane.

\citet[][; E14 hereafter]{2014MNRAS.439.3931E} were the first to perform systematic model fitting of the time-lags between the $0.3-1$ and $1.5-4\,\mathrm{keV}$ bands (henceforth, the soft excess vs continuum time-lags) for 12 AGN. They assumed the X-ray source to be point-like and located above the BH \citep[the so-called lamp-post geometry; e.g.][]{1991A&A...247...25M}, and calculated the response of the disc taking all relativistic effects into account. They deduced that the average X-ray source height is $\sim4r_{\mathrm{g}}$ with little scatter. \citet{2014MNRAS.438.2980C} were the first to model the time-lags between the $5-6\,\mathrm{keV}$ (which contains most of the photons from the red wing of a relativistically broadened Fe K$\alpha$ line) and $2-3\,\mathrm{keV}$ bands in the AGN NGC 4151. They used a similar procedure to E14, and deduced that the X-ray source height is $\sim7r_{\mathrm{g}}$, while the viewing angle of the system is $<30^{\circ}$. More recently, \citet[][; CY15, hereafter]{2015MNRAS.452..333C} simultaneously fitted, for the first time, the $4-6.5$ vs $2.5-4\,\mathrm{keV}$ time-lags and the $2-10\,\mathrm{keV}$ spectrum of Mrk 335. They found that the X-ray source is located very close to the central BH, at a height of $\sim2r_{\mathrm{g}}$.

Our main aim is to study the iron line vs continuum time-lag spectra (hereafter, the iron line vs continuum time-lags), within the context of the lamp-post geometry, similarly to E14, C14, and CY15. To this end, we chose the $5-7\,\rm{keV}$ band as representative of the energy band where most of the iron line photons will be (henceforth, the iron line band), and the $2-4\,\rm{keV}$ band as the energy band where the primary X-ray continuum dominates (henceforth, the continuum band). In our case, the exact choice of these two energy bands is relatively unimportant since, contrary to previous works (with the exception of CY15), we take into account the full disc reflection spectrum in both the iron line and continuum bands when constructing the theoretical lamp-post time-lag models, which we subsequently fitted to the observed iron line vs continuum time-lag spectra.

Our sample consists of seven AGN. We chose these objects because they are X-ray bright and have been observed many times by \textit{XMM-Newton}. We used all the existing \textit{XMM-Newton} archival data for these objects to estimate their iron line vs continuum time-lags. Our work improves significantly on the estimation of time-lags. We relied on \citet[][; EP16, hereafter]{2016A&A...591A.113E} to calculate time-lag estimates that are minimally biased, have known errors, and are approximately distributed as Gaussian variables. These properties render them appropriate for model fitting using traditional $\chi^2$ minimisation techniques.

Our results indicate that the data are consistent with a reverberation scenario, although the quality of the data is not high enough to estimate the various model parameters with high accuracy, except for the X-ray source height.

\section{Observations and data reduction} \label{sec2}

Table\,\ref{table1} lists the details of the \textit{XMM-Newton} observations we used. Columns\,1--4 show the source name, mass of the central BH in units of $10^6\,\mathrm{M}_{\odot}$, identification number (ID) of each observation, and net exposure in units of $\mathrm{ks}$, respectively.

\begin{table*}
\caption{\textit{XMM-Newton} observations log. Sources are listed in order of decreasing net exposure.}
\label{table1}      
\centering          
\begin{tabular}{c c c c c c c c}     
\hline\hline       
(1) & (2) & (3) & (4) & (1) & (2) & (3) & (4) \\
Source & $M_{\mathrm{BH}}$ & Obs. ID & Exp. & Source & $M_{\mathrm{BH}}$ & Obs. ID & Exp. \\
 & $(10^6\,\mathrm{M}_{\odot})$ & & (ks) & & $(10^6\,\mathrm{M}_{\odot})$ & & (ks) \\
\hline
\underline{1H 0707--495} & $2.3\pm0.7\,^a$ & & & \underline{NGC 4051} & $1.7^{+0.6}_{-0.5}\,^d$ \\
 & & 110890201 & 40.7 & & & 109141401 & 103.0\\
 & & 148010301 & 78.1 & & & 157560101 & 50.0 \\
 & & 506200201 & 38.7 & & & 606320101 & 45.3 \\
 & & 506200301 & 38.7 & & & 606320201 & 42.0 \\
 & & 506200401 & 40.6 & & & 606320301 & 21.1 \\
 & & 506200501 & 40.9 & & & 606320401 & 18.9 \\
 & & 511580101 & 121.6 & & & 606321301 & 30.2 \\
 & & 511580201 & 102.1 & & & 606321501 & 34.0 \\
 & & 511580301 & 104.1 & & & 606321601 & 41.5 \\
 & & 511580401 & 101.8 & & & 606321701 & 38.4 \\
 & & 554710801 & 96.1 & & & 606321801 & 18.8 \\
 & & 653510301 & 113.8 & & & 606322001 & 22.1 \\
 & & 653510401 & 125.7 & & & 606322101 & 29.2 \\
 & & 653510501 & 116.9 & & & 606322201 & 30.8 \\
 & & 653510601 & 119.5 & & & 606322301 & 42.3 \\
\underline{MCG--6-30-15} & $5.1^{+3.8}_{-2.4}\,^b$ & & & \underline{Ark 564} & $2.3\pm0.1\,^e$ \\
 & & 693781201 & 127.2 & & & 206400101 & 98.7 \\
 & & 693781301 & 129.7 & & & 670130201 & 59.1 \\
 & & 693781401 & 48.5 & & & 670130301 & 55.5 \\
 & & 111570101 & 33.1 & & & 670130401 & 56.7 \\
 & & 111570201 & 53.0 & & & 670130501 & 66.9 \\
 & & 029740101 & 80.6 & & & 670130601 & 57.0 \\
 & & 029740701 & 122.5 & & & 670130701 & 43.5 \\
 & & 029740801 & 124.1 & & & 670130801 & 57.8 \\
\underline{Mrk 766} & $1.8^{+1.6}_{-1.4}\,^c$ & & & & & 670130901 & 55.5 \\
 & & 109141301 & 116.9 & & & 006810101 & 10.6 \\
 & & 304030101 & 95.1 & \underline{NGC 7314} & $0.8\pm0.1\,^f$ \\
 & & 304030301 & 98.5 & & & 0111790101 & 43.3 \\
 & & 304030401 & 93.0 & & & 0311190101 & 77.5 \\
 & & 304030501 & 74.7 & & & 0725200101 & 124.7 \\
 & & 304030601 & 85.2 & & & 0725200301 & 130.6 \\
 & & 304030701 & 29.1 & \underline{Mrk 335} & $28\pm6\,^g$ \\
 & &           &      & & & 101040101 & 31.6 \\
 & &           &      & & & 306870101 & 122.5 \\
 & &           &      & & & 510010701 & 16.8 \\
 & &           &      & & & 600540501 & 36.9 \\
 & &           &      & & & 600540601 & 112.3 \\
\hline                  
\end{tabular}
\tablefoot{ \\
\tablefoottext{a}{\citet{2005ApJ...618L..83Z}} \\
\tablefoottext{b}{\citet{2005MNRAS.359.1469M}} \\
\tablefoottext{c}{\citet{2009ApJ...705..199B}} \\
\tablefoottext{d}{\citet{2010ApJ...721..715D}} \\
\tablefoottext{e}{Estimated using Eq.\,5 in \citet{2006ApJ...641..689V} for the $\mathrm{FWHM}(\mathrm{H}\beta)$ and $\lambda L_\lambda(5100\,\AA)$ values reported by \citet{2004ApJ...602..635R}} \\
\tablefoottext{f}{Estimated using Eq.\,3 in \citet{2009ApJ...698..198G} for the velocity dispersion value reported by \citet{2004ApJ...605..105C}} \\
\tablefoottext{g}{\citet{2012ApJ...744L...4G}} \\
}
\end{table*}

We processed data from the \textit{XMM-Newton} satellite using the Scientific Analysis System \citep[SAS, v. 14.0.0;][]{2004ASPC..314..759G}. We only used EPIC-pn \citep[][]{2001A&A...365L..18S} data. Source and background light curves were extracted from circular regions on the CCD, with the former having a fixed radius of 800 pixels ($40^{\prime\prime}$) centred on the source coordinates listed on the NASA/IPAC Extragalactic Database. The positions and radii of the background regions were determined by placing them sufficiently far from the location of the source, while remaining within the boundaries of the same CCD chip.

The source and background light curves were extracted in the iron line and continuum bands with a bin size of $100\,\mathrm{sec}$, using the SAS command evselect. We included the criteria PATTERN==0--4 and FLAG==0 in the extraction process, which select only single- and double-pixel events and reject bad pixels from the edges of the detector CCD chips. Periods of high solar flaring background activity were determined by observing the $10-12\,\mathrm{keV}$ light curves (which contain very few source photons) extracted from the whole surface of the detector, and subsequently excluded during the source and background light curve extraction process.

We checked all source light curves for pile-up using the SAS task epatplot and found that only observations 670130201, 670130501, and 670130901 of Ark 564 are affected. For those observations we used annular instead of circular source regions with inner radii of 280, 200, and 250 pixels (the outer radii were held at 800 pixels), respectively, which we found to adequately reduce the effects of pile-up.

The background light curves were then subtracted from the corresponding source light curves using the SAS command epiclccorr. Most of the resulting light curves were continuously sampled, except for a few cases that contained a small ($\lesssim5\%$ of the total number of points in the light curve) number of missing points. These were either randomly distributed throughout the duration of an observation, or appeared in groups of $\lesssim10$ points. We replaced the missing points by linear interpolation, with the addition of the appropriate Poisson noise.

\section{Time-lag estimation} \label{sec3}

\begin{table*}[ht!]
\caption{\textit{XMM-Newton} light curve characteristics relevant to the estimation of time-lags}
\label{table2}
\centering          
\begin{tabular}{c c c c c c}
\hline\hline
Source & Segment duration & No. of segments & Mean count rate & Mean count rate & Max. frequency \\
 & (ks) & $m$ & (cts/sec; $5-7\,\mathrm{keV}$) & (cts/sec; $2-4\,\mathrm{keV}$) & $\nu_{\mathrm{max}}$ \\
\hline
1H 0707--495 & 20.2 & 57 & 0.02 & 0.10 & $6.9\times10^{-4}$ \\
MCG--6-30-15 & 24.2 & 28 & 0.85 & 3.12 & $1.0\times10^{-3}$ \\
Mrk 766 & 23.2 & 24 & 0.30 & 1.22 & $6.9\times10^{-4}$ \\
NGC 4051 & 20.6 & 22 & 0.35 & 1.02 & $1.2\times10^{-3}$ \\
Ark 564 & 27.7 & 18 & 0.34 & 2.15 & $7.2\times10^{-4}$ \\
NGC 7314 & 20.7 & 17 & 0.59 & 1.88 & $1.1\times10^{-3}$ \\
Mrk 335 & 26.9 & 12 & 0.23 & 0.93 & $3.0\times10^{-4}$ \\
\hline
\end{tabular}
\vspace{1cm}
\end{table*}
   
We used standard Fourier techniques \citep[see, e.g.,][]{1999ApJ...510..874N,2014A&ARv..22...72U} to estimate time-lags between light curves in the iron line and continuum bands for our sample.

We denote by $\{x(t_r),y(t_r)\}$ a pair of light curves in two energy bands, where $t_r=\Delta t,2\Delta t,\ldots,N\Delta t$, $N$ is the number of points and $\Delta t=100\,\mathrm{sec}$ is the time bin size. The discrete Fourier transforms (DFTs), $\{\zeta_x(\nu_p),\zeta_y(\nu_p)\}$, of the light curves are
\noindent
\begin{align}
\label{eq1}
\zeta_x(\nu_p) &\equiv\sqrt{\frac{\Delta t}{N}}\sum_{r=1}^{N}[x(t_r)-\overline{x}]\mathrm{e}^{-\mathrm{i}2\pi\nu_pt_r}, \\
\label{eq2}
\zeta_y(\nu_p) &\equiv\sqrt{\frac{\Delta t}{N}}\sum_{r=1}^{N}[y(t_r)-\overline{y}]\mathrm{e}^{-\mathrm{i}2\pi\nu_pt_r},
\end{align}
\noindent
where $\overline{x}$ and $\overline{y}$ are the light-curve sample means, and $\nu_p=p/N\Delta t$ ($p=1,2,\ldots,N/2$). The cross-periodogram, $I_{xy}(\nu_p)$, of the light-curve pair is defined as \citep[][; henceforth P81]{Priestley:81}
\noindent
\begin{equation} \label{eq3}
I_{xy}(\nu_p)\equiv\zeta_x(\nu_p)\zeta^{*}_y(\nu_p).
\end{equation}
\noindent
The cross-periodogram is an estimator of the intrinsic cross-spectrum (CS), $C_{xy}(\nu_p)$, which is a measure of the cross-correlation between two random signals in Fourier space. The cross-periodogram is generally biased, in the sense that the mean of $I_{xy}(\nu_p)$ is not equal to $C_{xy}(\nu_p)$. The traditional time-lag estimator, which we define below, is based on the cross-periodogram. Therefore, the statistical properties of the two estimators are closely dependent. As shown by EP16, there are two main factors that contribute to the bias of the cross-periodogram: the finite duration of the light curves, and their sampling rate and time bin size (in our work, the sampling rate is equal to the time bin size).

Discrete sampling of a continuous process introduces aliasing effects to the CS of the resulting discrete process, which is only defined in the frequency range $[-1/2\Delta t,1/2\Delta t]$ and is equal to the superposition of the intrinsic CS at frequencies $\nu,\nu\pm1/\Delta t,\nu\pm2/\Delta t,$ etc. Aliasing effects are reduced when the light curves are binned. They are similar to the aliasing effects in the power-spectral density (PSD) of a light curve, although while PSDs are always positive, this is generally not the case with CS. As a result, aliasing effects are more complex in this case. EP16 found that light-curve binning generally causes the measured time-lags to converge to zero at frequencies $\gtrsim\nu_{\mathrm{Nyq}}/2$, where $\nu_{\mathrm{Nyq}}=1/2\Delta t$ is the Nyquist frequency. In this work $\nu_{\mathrm{Nyq}}=5\times10^{-3}\,\mathrm{Hz}$, and hence we only computed cross-periodograms at frequencies $\le2.5\times10^{-3}\,\mathrm{Hz}$.

Owing to the finite light-curve duration, the mean of the cross-periodogram is equal to the convolution of the intrinsic CS (as modified by aliasing effects) with a particular window function, just like the case of the periodogram \citep[i.e. the traditional PSD estimator; see e.g.][]{1993MNRAS.261..612P}. However, the effects of this convolution on the time-lag estimates cannot be predicted a priori, since they depend on the shape of the (unknown) intrinsic CS (and not just on the intrinsic time-lag spectrum). They were quantitatively investigated by EP16, who considered three different types of time-lag spectra that are typically observed between X-ray light curves of accreting systems: constant time-lags, time-lags with a power-law dependence on frequency, and time-lags that have a characteristic oscillatory behaviour with frequency, similar to what is expected in a reverberation scenario. For the model CS they considered, they concluded that time-lag estimates based on the cross-periodogram will not be significantly biased, in the sense that their mean will be within $\sim15\%$ (in absolute value) of their corresponding intrinsic values when the light-curve duration is $\gtrsim20\,\mathrm{ks}$.

The cross-periodogram has a large and unknown variance. As a result, this feature will be shared by the time-lag estimates computed from it. This problem is ameliorated in practice by either binning together $m$ neighbouring frequency ordinates of the cross-periodogram (a process called smoothing), and/or binning different cross-periodogram ordinates at a given frequency obtained from $m$ distinct light-curve pairs. If, as is often the case in practice, the real and imaginary parts of the intrinsic CS vary in a non-linear fashion over the smoothed frequency range, then smoothing will introduce an additional source of bias to the cross-periodogram. This bias can only be taken into account a posteriori when fitting observed time-lags by prescribing a model CS (and not just a model time-lag spectrum), as it affects the cross-periodogram itself.

Since this is a complicated model-dependent procedure, we did not perform any smoothing on the cross-periodograms. We instead divided the available \textit{XMM-Newton} observations of each source into shorter segments of duration $20-40\,\mathrm{ks}$. The segment duration for each source (listed in Col.\,2 of Table\,\ref{table2}) was determined in such a way as to maximise their number, $m$, for the total available light curves ($m$ is listed in Col.\,3 of Table\,\ref{table2}).

   \begin{figure}
   \centering
   \includegraphics[width=\hsize]{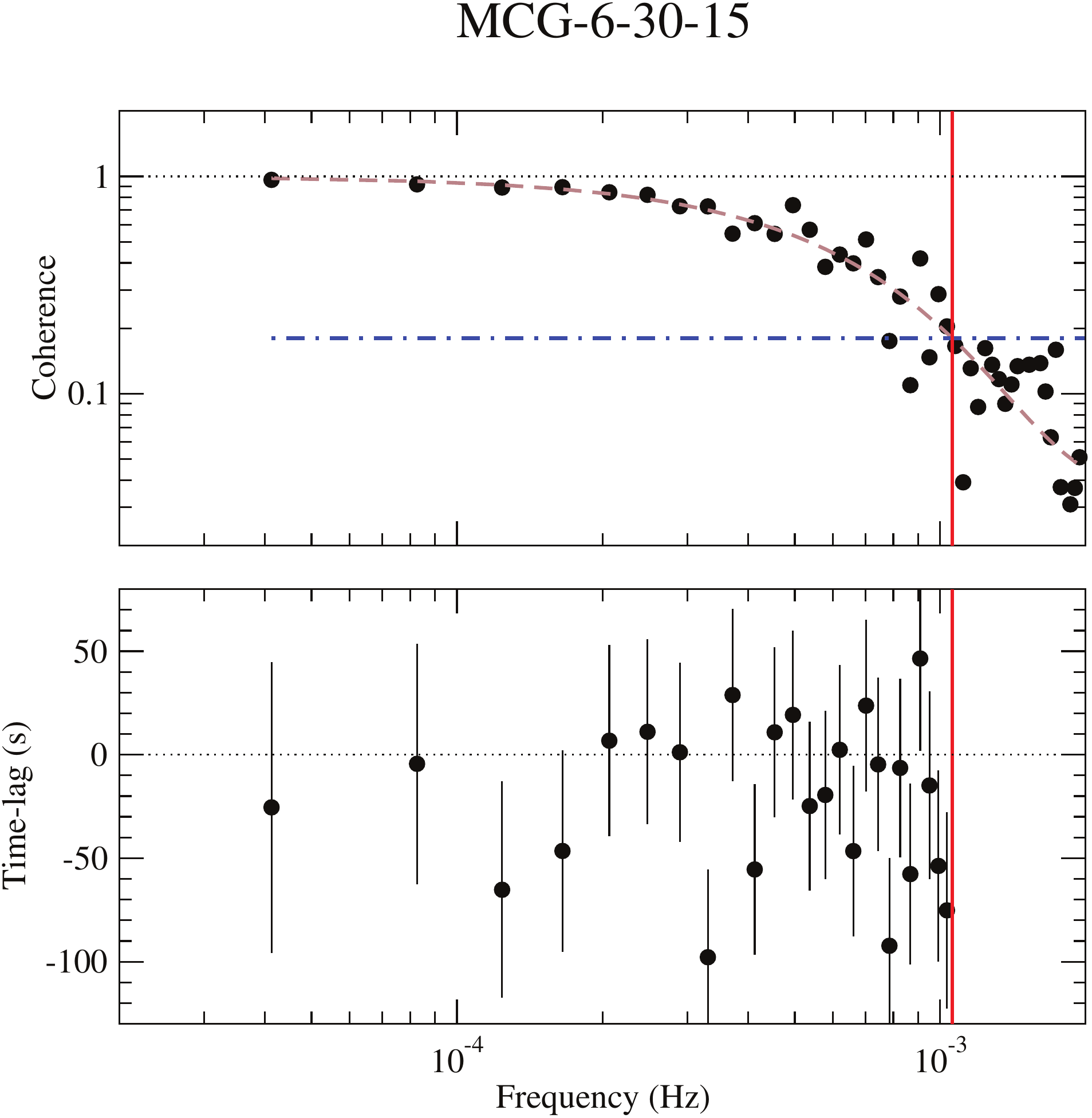}
      \caption{Sample iron line vs continuum coherence function (top panel) and time-lag spectrum (bottom panel) of MGG--60-30-15, estimated using the data listed in Table\,\ref{table2}. The dashed brown line in the top panel shows the best-fit model to the sample coherence. The continuous red vertical line indicates the highest frequency up to which time-lags should be estimated, and the horizontal blue dotted-dashed line indicates the coherence value at this frequency (see Sect.\,\ref{sec3}).}
         \label{fig1}
   \end{figure}
   
   \begin{figure}[h!]
   \centering
   \includegraphics[width=\hsize]{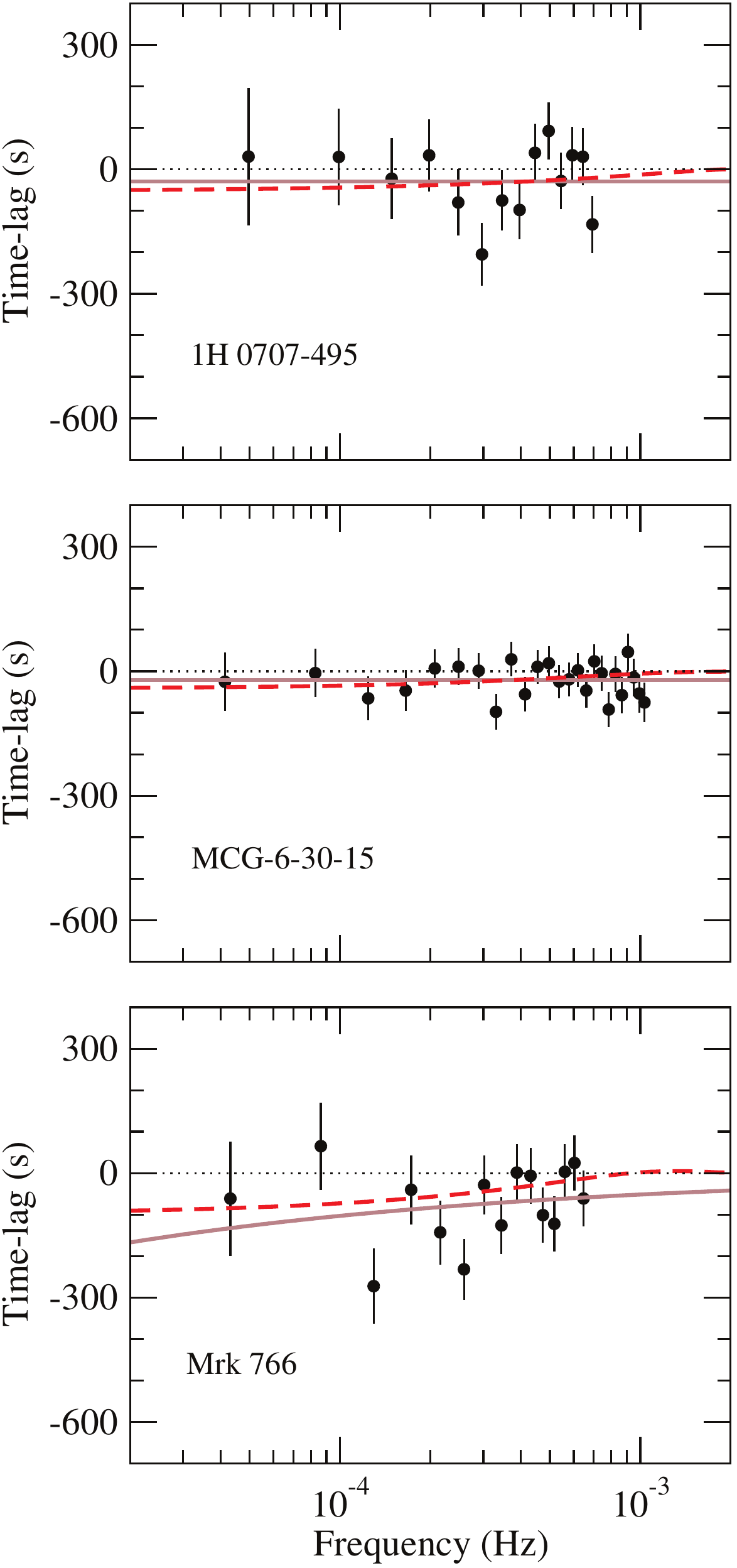}
      \caption{Observed iron line vs continuum time-lag spectra for 1H 0707--495 (first row), MCG--6-30-15 (second row), and Mrk 766 (third row). The solid brown and dashed red lines indicate the best-fit models A and B, respectively, to each time-lag spectrum (see Sect.\,\ref{sec5} for details on these models).}
         \label{fig2}
   \end{figure}
   
   \begin{figure}[h!]
   \centering
   \includegraphics[width=\hsize]{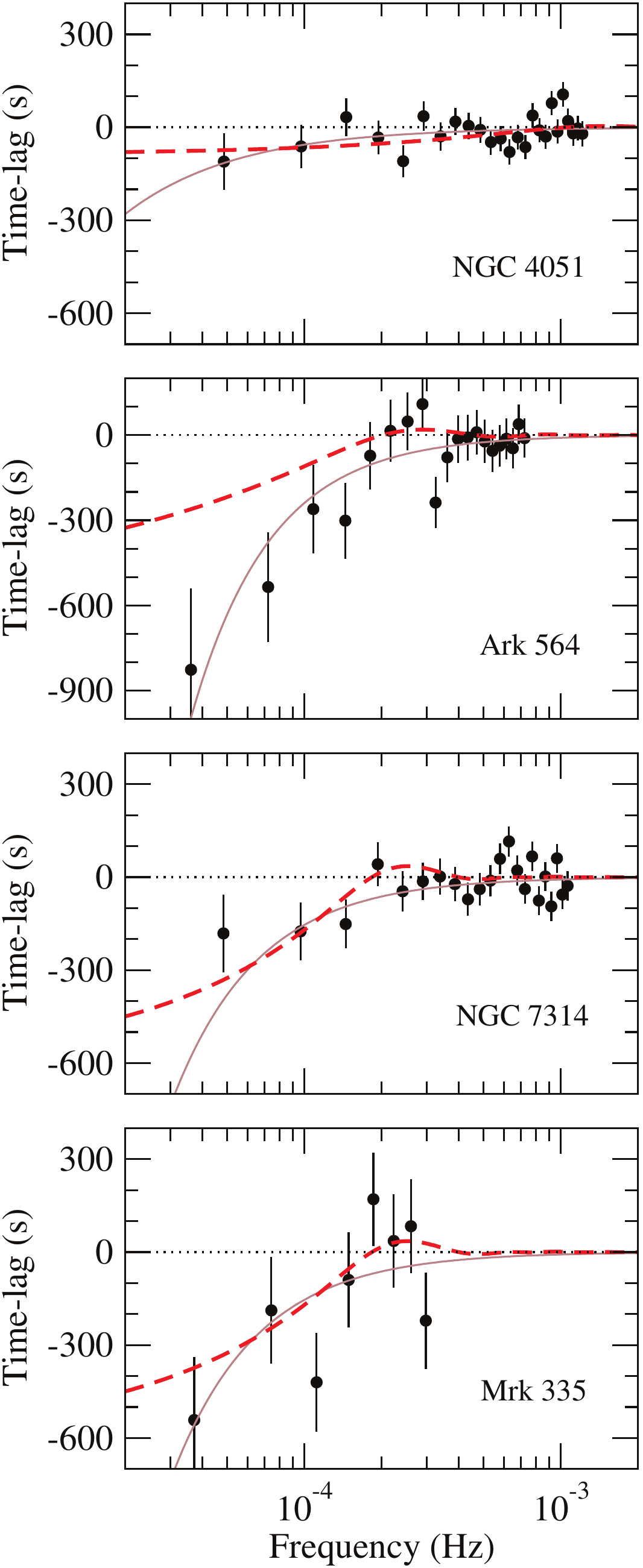}
      \caption{Same as in Fig.\,\ref{fig2} for NGC 4051 (first row), Ark 564 (second row), NGC 7314 (third row), and Mrk 335 (fourth row).}
         \label{fig3}
   \end{figure}
   
For each segment we calculated the cross-periodogram according to Eq.\,\ref{eq3}, and adopted
\noindent
\begin{equation} \label{eq4}
\hat{C}_{xy}(\nu_p)=\frac{1}{m}\sum_{k=1}^{m}I^{(k)}_{xy}(\nu_p)
\end{equation}
\noindent
and
\noindent
\begin{equation} \label{eq5}
\hat{\tau}_{xy}(\nu_p)\equiv\frac{1}{2\pi\nu_p}\mathrm{arg}[\hat{C}_{xy}(\nu_p)]
\end{equation}
\noindent
as our estimates of the CS and time-lag spectrum, respectively ($I^{(k)}_{xy}(\nu_p)$ is the cross-periodogram of the $k-$th segment at frequency $\nu_p$). We adopted the standard convention of defining $\mathrm{arg}[\hat{C}_{xy}(\nu_p)]$ on the interval $(-\pi,\pi]$. The analytic error estimate of $\hat{\tau}_{xy}(\nu_p)$ is given by \citep[e.g. P81;][]{1999ApJ...510..874N}
\noindent
\begin{equation} \label{eq6}
\sigma_{\hat{\tau}}(\nu_p)\equiv\frac{1}{2\pi\nu_p}\frac{1}{\sqrt{2m}}\sqrt{\frac{1-\hat{\gamma}^2_{xy}(\nu_p)}{\hat{\gamma}^2_{xy}(\nu_p)}},
\end{equation}
\noindent
where \citep[e.g. P81;][]{1997ApJ...474L..43V}
\noindent
\begin{equation} \label{eq7}
\hat{\gamma}^2_{xy}(\nu_p)\equiv\frac{|\hat{C}_{xy}(\nu_p)|^2}{\hat{P}_x(\nu_p)\hat{P}_y(\nu_p)}.
\end{equation}
\noindent
$\hat{P}_x(\nu_p)$ and $\hat{P}_y(\nu_p)$ are the traditional periodograms of the two light curves, which are also calculated by binning over $m$ segments. Equation\,\ref{eq7} defines an estimator of the so-called coherence function. This function is defined on the interval $[0,1]$ and quantifies the degree of linear correlation between sinusoidal components of two light curves at each frequency.

Figure\,\ref{fig1} shows the sample iron line vs continuum coherence and time-lag spectrum of MCG--6-30-15 (top and bottom panel, respectively), which were calculated using Eqs.\,\ref{eq7} and \ref{eq5}. The sample coherence decreases to zero with increasing frequency. This loss of coherence is mostly caused by Poisson noise. In the presence of measurement errors, even if the intrinsic coherence is unity at all frequencies, the resulting coherence will decrease towards zero at frequencies where the amplitude of experimental noise variations dominates the amplitude of the intrinsic variations. The sample coherence will, however, always converge to a constant value $1/m$ at these frequencies. EP16 found that this decrease can be reasonably approximated by an exponential function of the form
\noindent
\begin{equation} \label{eq8}
\hat\gamma^2_{xy}(\nu)=\left(1-\frac{1}{m}\right)\mathrm{exp}[-(\nu/\nu_0)^q]+\frac{1}{m},
\end{equation}
\noindent
where $\nu_0$ and $q$ are parameters that are determined by fitting this function to the coherence estimates. This was empirically found by EP16 to fit the sample coherence well, using many simulations of light curves in the case of various model CS and light curve signal-to-noise ratios (S/N). An example of such a fit to the coherence estimates of MCG--6-30-15 is shown in the top panel of Fig.\,\ref{fig1} (brown dashed line). The fit describes the sample coherence function well (this was the case for all sources).

According to Eq.\,\ref{eq6}, the error of the time-lag estimates increases as the coherence decreases. Therefore, above a certain maximum frequency, $\nu_{\mathrm{max}}$, when the coherence is sufficiently small (i.e. $\sim0$), we expect that Poisson noise will severely affect the reliability of the time-lag estimates. The effects of Poisson noise on the bias and distributions of the time-lag estimates were quantitatively investigated by EP16. They found that $\nu_{\mathrm{max}}$ decreases as the S/N of the light curves decreases. In addition, $\nu_{\mathrm{max}}$ is mainly affected by the energy band with the lowest mean count rate, which in our case corresponds to the iron line band. In Cols.\,4 and 5 of Table\,\ref{table2} we list the mean count rate in the iron line and continuum band, respectively. According to EP16, $\nu_{\mathrm{max}}$ corresponds approximately to the frequency at which the sample coherence function becomes equal to $1.2/(1+0.2m)$. Above $\nu_{\mathrm{max}}$, EP16 found that Poisson noise has the following effects on the time-lag estimates: (a) The analytic error estimate given by Eq.\,\ref{eq6} increasingly underestimates their true scatter, and (b) their distribution becomes uniform and symmetrical about a zero time-lag value. As a result, the time-lag estimates become biased, in the sense that their mean converges to zero, independent of the intrinsic time-lag spectrum. Below $\nu_{\mathrm{max}}$, and as long as $m\gtrsim10$, the mean of the time-lag estimates is not affected, Eq.\,\ref{eq6} provides a reliable estimate of their true scatter, and their distribution is approximately Gaussian.

We therefore fitted the coherence estimates of each source to the exponential function given by Eq.\,\ref{eq8} (as we did for MCG--6-30-15), and equated this function to the constant $1.2/(1+0.2m)$ to estimate $\nu_{\mathrm{max}}$ in each case. The values of $\nu_{\mathrm{max}}$ calculated thus are listed in Col.\,6 of Table\,\ref{table2}. We did not estimate time-lags above this frequency. Instead of using the values of the sample coherence function to determine the errors of the time-lag estimates according to Eq.\,\ref{eq6}, we used the values of the best-fit exponential model. We found that the resulting errors are more representative of the observed scatter of the time-lag estimates, although the differences are small ($\lesssim20\%$). The iron line vs continuum time-lag estimates for each source, along with their errors, obtained by the above procedure are shown in Figs.\,\ref{fig2} and \ref{fig3}.

\section{Theoretical modelling of the time-lag spectra} \label{sec4}

In this section we describe the basic physical and geometrical properties of the lamp-post model and show how we determined the corresponding theoretical iron line vs continuum time-lag spectra. All physical quantities in the lamp-post model are estimated in geometrised units ($G=c=1$) and scale with $M_{\mathrm{BH}}$. Thus, for instance, time-scales have to be multiplied by a factor $t_{\mathrm{g}}\equiv GM_{\mathrm{BH}}/c^3\sim5(M_{\mathrm{BH}}/10^6\,\mathrm{M}_{\odot})\,\mathrm{sec}$ to be converted into units of seconds.

\subsection{Geometrical layout of the model} \label{subsec41}

The lamp-post model consists of a BH, surrounded by an equatorial accretion disc, that is illuminated by an X-ray source located on the disc symmetry axis. The parameters of the model are the mass and spin ($a$) of the BH, the height ($h$) of the X-ray source, and the viewing angle ($\theta$) of a distant observer with respect to the disc axis.

The disc is assumed to be geometrically thin and Keplerian, co-rotating with the BH, with a radial extent ranging from the innermost stable circular orbit (ISCO), $r_{\mathrm{ISCO}}$, up to an outer radius $r_{\mathrm{out}}$. The BH spin uniquely defines $r_{\mathrm{ISCO}}$. When measured in geometrised units, the spin can attain any value between zero and unity, with $a=0$ ($r_{\mathrm{ISCO}}=6r_{\mathrm{g}}$) and $a=1$ ($r_{\mathrm{ISCO}}=1r_{\mathrm{g}}$) indicating a non-spinning (i.e. Schwarzschild) and maximally spinning (i.e. extreme Kerr) BH, respectively. The X-ray source is assumed to be point-like and located at a fixed position above the BH. It emits isotropically with an intrinsic (i.e. rest-frame) spectrum of $\mathscr{N}(t)E^{-2}\mathrm{exp}(-E/300\,\mathrm{keV})$. We assumed it to be variable in amplitude only, and that $\mathscr{N}(t)$ is a stationary random process (i.e. that it has a finite and time-independent mean and variance).

\subsection{Observed fluxes at infinity} \label{subsec42}

We assumed that the total flux recorded by an observer at a very large distance in a given energy band $\mathcal{E}=[E_1(\mathrm{keV}),E_2(\mathrm{keV})]$ is $\mathscr{F}_{\mathcal{E}}(t;a,\theta,h,r_{\mathrm{out}})$. This flux is equal to the sum of the continuum and reprocessed flux from the disc, $\mathscr{F}^{(\mathrm{c})}_{\mathcal{E}}(t;a,\theta,h)$ and $\mathscr{F}^{(\mathrm{r})}_{\mathcal{E}}(t;a,\theta,h,r_{\mathrm{out}})$, respectively. In other words,
\noindent
\begin{align} \label{eq9}
\nonumber
\mathscr{F}_{\mathcal{E}}(t;a,\theta,h)= & \mathscr{F}^{(\mathrm{c})}_{\mathcal{E}}(t;a,\theta,h)+\mathscr{F}^{(\mathrm{r})}_{\mathcal{E}}(t;a,\theta,h,r_{\mathrm{out}}) \\
\nonumber
= & \mathscr{F}^{(\mathrm{c})}_{\mathcal{E}}(t;a,\theta,h) \\
& +\int_{-\infty}^{\infty}\Psi_{\mathcal{E}}(t';a,\theta,h,r_{\mathrm{out}})\mathscr{F}^{(\mathrm{c})}_{\mathcal{E}}(t-t';a,\theta,h)\mathrm{d}t',
\end{align}
\noindent
where $\Psi_{\mathcal{E}}(t';a,\theta,h,r_{\mathrm{out}})$ is the so-called response function, which quantifies the response of the disc to an instantaneous flare of continuum emission. We define the normalisation of the response function such that its time-integrated value is equal to the observed ratio of reprocessed-to-continuum photons.

The observed continuum spectrum differs in amplitude from its rest-frame value as a result of relativistic effects \citep[][]{2011ApJ...731...75D}. This is quantified by the factor $\mathscr{G}(a,\theta,h)$, such that $\mathscr{F}^{(\mathrm{c})}_{\mathcal{E}}(t;a,\theta,h)=\mathscr{N}_{\mathcal{E}}(t)\mathscr{G}(a,\theta,h)\int_{E_1}^{E_2}E^{-2}\mathrm{exp}(-E/300\,\mathrm{keV})\mathrm{d}E$. The dependence of the various terms on the right-hand side of Eq.\,\ref{eq9} on the parameters of the lamp-post model were explicitly included and are henceforth be omitted for reasons of brevity.

\subsection{Time-lag spectra} \label{subsec43}

We assumed that the total photon fluxes observed in the iron line and continuum bands are $\mathscr{F}_{5-7}(t)$ and $\mathscr{F}_{2-4}(t)$, respectively. According to Eq.\,\ref{eq9},
\noindent
\begin{align}
\label{eq10}
\mathscr{F}_{5-7}(t) &=\mathscr{F}^{(\mathrm{c})}_{5-7}(t)+\int_{-\infty}^{\infty}\Psi_{5-7}(t')\mathscr{F}^{(\mathrm{c})}_{5-7}(t-t')\mathrm{d}t', \\
\label{eq11}
\mathscr{F}_{2-4}(t) &=\mathscr{F}^{(\mathrm{c})}_{2-4}(t)+\int_{-\infty}^{\infty}\Psi_{2-4}(t')\mathscr{F}^{(\mathrm{c})}_{2-4}(t-t')\mathrm{d}t'.
\end{align}
\noindent
The cross-correlation function (CCF) between the iron line and continuum bands, $R_{5-7,2-4}(\tau)$, is then
\noindent
\begin{equation} \label{eq12}
R_{5-7,2-4}(\tau)\equiv\mathrm{E}\{[\mathscr{F}_{2-4}(t)-\mu_{2-4}][\mathscr{F}_{5-7}(t+\tau)-\mu_{5-7}]\},
\end{equation}
\noindent
where $\mathrm{E}$ is the expectation operator, and $\mu_{5-7}$ ($\mu_{2-4}$) is the mean flux in the iron line (continuum) band. The CS, $C_{5-7,2-4}(\nu)$, between the two energy bands is, by definition, the Fourier transform of the CCF. Hence (see Appendix \ref{appa} for a more detailed derivation)
\noindent
\begin{align} \label{eq13}
\nonumber
C_{5-7,2-4}(\nu) &\equiv\int_{-\infty}^{\infty}R_{5-7,2-4}(\tau)\mathrm{e}^{-\mathrm{i}2\pi\nu\tau}\mathrm{d}\tau \\
&=C^{(\mathrm{c})}_{5-7,2-4}(\nu)[1+\tilde{\Psi}_{5-7}(\nu)][1+\tilde{\Psi}_{2-4}(\nu)]^{*},
\end{align}
\noindent
where the asterisk denotes complex conjugation, $\tilde{\Psi}_{\mathcal{E}}(\nu)\equiv\int_{-\infty}^{\infty}\Psi_{\mathcal{E}}(t)\mathrm{e}^{-\mathrm{i}2\pi\nu t}\mathrm{d}t$ is the Fourier transform of the response function, and $C^{(\mathrm{c})}_{5-7,2-4}(\nu)$ is the CS of the continuum emission. 

The iron line vs continuum time-lag spectrum, $\tau_{5-7,2-4}(\nu)$, is defined as $\tau_{5-7,2-4}(\nu)\equiv(2\pi\nu)^{-1}\mathrm{arg}[C_{5-7,2-4}(\nu)]$. Given our adopted convention, a positive value of $\tau_{5-7,2-4}(\nu)$ indicates that variations in the iron line band lead variations in the continuum band (and vice versa). According to Eq.\,\ref{eq13},
\noindent
\begin{equation} \label{eq14}
\tau_{5-7,2-4}(\nu)=\tau^{(\mathrm{c})}_{5-7,2-4}(\nu)+\tau^{(\mathrm{r})}_{5-7,2-4}(\nu).
\end{equation}
\noindent
The equation above shows that the time-lags between the observed variations in the two energy bands equal the sum of two terms; time-lags between variations in the X-ray continuum, $\tau^{(\mathrm{c})}_{5-7,2-4}(\nu)$, and time-lags due to reprocessed X-ray emission from the disc, $\tau^{(\mathrm{r})}_{5-7,2-4}(\nu)$ (henceforth, the continuum and reverberation time-lags, respectively). The continuum time-lags are given by $\tau^{(\mathrm{c})}_{5-7,2-4}(\nu)\equiv(2\pi\nu)^{-1}\mathrm{arg}[C^{(\mathrm{c})}_{5-7,2-4}(\nu)]$, while the reverberation time-lags are given by
\noindent
\begin{equation} \label{eq15}
\tau^{(\mathrm{r})}_{5-7,2-4}(\nu)\equiv\frac{1}{2\pi\nu}\mathrm{arg}\{[1+\tilde{\Psi}_{5-7}(\nu)][1+\tilde{\Psi}_{2-4}(\nu)]^{*}\}.
\end{equation}
\noindent
This function is uniquely determined by the disc response functions in the iron line and continuum bands.

\subsection{Response functions in the lamp-post geometry} \label{subsec44}

To determine the response function of the disc, we assumed that the primary X-ray source isotropically emits a flare of duration equal to $1t_{\mathrm{g}}$. Upon being illuminated, each area element of the disc responds to this flare by isotropically and instantaneously emitting a reflection spectrum in its rest-frame. We assumed that the reprocessed flux is proportional to the incident flux and that the disc material is neutral, with an iron abundance equal to the solar value. We then used the rest-frame reflection spectrum computed with the multi-scattering code NOAR \citep{2000A&A...357..823D}. We determined the time-varying $0.1-8\,\mathrm{keV}$ disc reflection spectrum at infinity, with a time resolution of $0.1t_{\mathrm{g}}$ and energy resolution of $20\,\mathrm{eV}$, taking all relativistic effects into account \citep[e.g. gravitational and Doppler energy shifts, light bending, and time delays;][]{2006AN....327..961K}.

Finally, we calculated the disc response function in the iron line and continuum bands by integrating the observed disc reflection spectrum in the appropriate energy ranges. In Appendix \ref{appb} we show how the disc response functions depend on the parameters of the lamp-post model, while in Appendix \ref{appc} we show how we computed the model time-lag spectra given by Eq.\,\ref{eq15} from the numerically computed disc response functions. In Appendix \ref{appd} we discuss how these model time-lag spectra depend on the parameters of the lamp-post model.

The response functions we computed are similar to those presented by \citet{1999ApJ...514..164R} and E14, although our approach is different. They calculated the response function considering only the Fe K$\alpha$ photons emitted in the disc rest-frame. In contrast, we counted all the photons from the reflection component that an observer will detect in an energy band (within $0.1-8\,\mathrm{keV}$) at each time step. We therefore considered the total reflection spectrum, as emitted by the disc rest frame, hence we computed the total reflection response function and self-consistently included the reflection component in both the continuum and iron line bands along with the X-ray continuum emission. This approach is be more appropriate for comparing our predictions with data. Our approach is similar to the one adopted by CY15, although we did not consider the effects of disc ionisation.

\section{Fitting procedure} \label{sec5}

As explained in Sect.\,\ref{sec3}, our time-lag estimates should be approximately distributed as Gaussian random variables. Fitting the observed iron line vs continuum time-lags was therefore based on minimising the $\chi^2$ function, which is defined as
\noindent
\begin{equation} \label{eq16}
\chi^2(a_1,a_2,\ldots,a_q)\equiv\sum_{p=1}^{n}\frac{[\hat{\tau}(\nu_p)-\tau(\nu_p;a_1,a_2,\ldots,a_q)]^2}{\sigma^2_{\hat{\tau}}(\nu_p)},
\end{equation}
\noindent
where $\{a_1,a_2,\ldots,a_q\}$ are the parameters of the model, $\hat{\tau}(\nu_p)$ is the time-lag estimate with error $\sigma_{\hat{\tau}}(\nu_p)$, $\tau(\nu_p;a_1,a_2,\ldots,a_q)$ is the model time-lag spectrum, and $n$ is the number of time-lag estimates. The location of the $\chi^2(a_1,a_2,\ldots,a_q)$ minimum, say $\chi^2_{\mathrm{min}}$, determines the best-fit parameter values. Their corresponding 68\% (95\%) confidence intervals are determined by the standard $\Delta\chi^2=1$ ($\Delta\chi^2=4$) method for one independent parameter. Unless otherwise mentioned, confidence intervals of best-fit parameters are henceforth quoted at the 68\% level.

As we showed in Sect.\,\ref{subsec43}, the observed time-lags should depend on both the continuum and reverberation time-lags. We thus considered two different model time-lag spectra, one for the continuum and the other for the reverberation time-lags. We describe them in more detail below.

\subsection{Model A: Continuum time-lags model} \label{subsec51}

In AGN and X-ray binaries, time-lag spectra between X-ray light curves are typically observed to have a power-law dependence on frequency. High-energy bands are delayed with respect to lower energy bands, and the magnitude of the time-lags decreases with increasing frequency, typically following a power-law like form \citep[e.g.][]{1989Natur.342..773M,1996MNRAS.280..227N,1999ApJ...510..874N,2001ApJ...554L.133P,2004MNRAS.348..783M,
2006MNRAS.372..401A,2008MNRAS.388..211A,2009ApJ...700.1042S}. In addition, the magnitude of these time-lags is observed to increase with increasing energy separation between the two energy bands. We therefore considered a power-law model of the form
\noindent
\begin{equation} \label{eq17}
\tau^{(\mathrm{c})}_{5-7,2-4}(\nu)=-A(\nu/10^{-4}\,\mathrm{Hz})^{-s},
\end{equation}
\noindent
where $A$ and $s$ are positive, to account for the continuum time-lags. These continuum time-lags are expected to be negative in our case, meaning that variations in the iron line band should be delayed with respect to variations in the continuum band.

\subsection{Model B: Reverberation time-lags model} \label{subsec52}

The model B time-lag spectrum corresponds to the function $\tau^{(\mathrm{r})}_{5-7,2-4}(\nu)$ given by Eq.\,\ref{eq15}, that is to say, it accounts for the reverberation time-lags. This function is uniquely determined by the Fourier transforms of the response functions in the iron line and continuum bands. Since these response functions are not given by an analytical formula, we had to numerically compute them (following the procedure outlined in Sect.\,\ref{subsec44}) on a grid of points corresponding to different combinations of $\{a,\theta,h,M_{\mathrm{BH}}\}$ values (we set $r_{\rm out}=10^3r_{\mathrm{g}}$ in all cases).

\section{Results} \label{sec6}

\begin{table*}[ht!]
\caption{Model fit results}             
\label{table3}      
\centering          
\begin{tabular}{c c c c c c c}
\hline\hline
 & \multicolumn{3}{c}{Model A} & \multicolumn{3}{c}{Model B} \\
Source & $A$ & $s$ & $\chi^2_{\mathrm{min}}/\mathrm{dof}$ & $h$ & $M_{\mathrm{BH}}\,^{a}$ & $\chi^2_{\mathrm{min}}/\mathrm{dof}$ \\
 & ($\mathrm{sec}$) &  &  & ($r_{\mathrm{g}}$) & ($10^6\,\mathrm{M}_{\odot}$) & $a$ free ($a=0$, $a=1$)$\,^{b}$ \\
\hline
1H 0707--495 & $33^{+46}_{-24}$ & $<0.9$ & $15.9/12$ & $<20$ & $2.3$ & $15.9/12$ (16.0, 15.9) \\
MCG--6-30-15 & $21^{+23}_{-9}$ & $<0.6$ & $19.0/23$ & $<3$ & $5.1$ & $18.4/23$ (20.7, 27.2) \\
Mrk 766 & $103\pm39$ & $<0.6$ & $18.6/13$ & $22^{+12}_{-10}$ & $1.8$ & $20.5/13$ (20.7, 20.5) \\
NGC 4051 & $56^{+24}_{-33}$ & $1.0^{+0.8}_{-0.5}$ & $28.3/23$ & $<30$ & $1.7$ & $29.3/23$ (29.3, 29.5) \\
Ark 564 & $239^{+57}_{-56}$ & $1.4^{+0.2}_{-0.3}$ & $13.5/18$ & $>28$ & $2.3$ & $23.2/18$ (23.2, 23.2) \\
NGC 7314 & $94\pm37$ & $1.3^{+0.5}_{-0.4}$ & $25.9/20$ & $>82$ & $0.8$ & $25.2/20$ (25.2, 25.2) \\
Mrk 335 & $154^{+73}_{-79}$ & $1.3^{+0.8}_{-0.5}$ & $8.3/6$ & $7^{+2}_{-3}$ & $28$ & $7.9/6$ (9.6, 7.9) \\
\hline                  
\end{tabular}
\tablefoot{ \\
\tablefoottext{a}{Fixed to the values listed in Col.\,2 of Table\,\ref{table1}.} \\
\tablefoottext{b}{Note that the number of degrees of freedom is increased by one when $a$ is fixed compared to the values listed for the case when it is a free parameter.}
}
\vspace{1cm}
\end{table*}

According to Eq.\,\ref{eq14}, we should fit the observed time-lags with the sum of models A and B. However, we discovered that due to the limited frequency range of the observed time-lag spectra and the relatively large errors of the time-lag estimates, it was not possible to simultaneously constrain the parameters of both models in a meaningful way. We therefore decided to fit the two models separately to the data and then investigate whether they provided a good fit or not. The only exceptions were Ark 564 and NGC 7314, whose observed time-lag spectra we also fitted to a combined model A+B as well for reasons we discuss in Sect.\,\ref{subsec62} below.

The continuum time-lags model (i.e. model A) is defined by Eq.\,\ref{eq17}. The model has two free parameters ($A$ and $s$). For each observed time-lag spectrum shown in Figs.\,\ref{fig2} and \ref{fig3}, we calculated $\chi^2(A,s)$ using Eq.\,\ref{eq16}. We then minimised this function numerically using the Levenberg-Marquardt method, and determined the best-fit values and confidence intervals of the model A parameters.

For the reverberation time-lags (i.e. model B), the parameter space we considered for the model parameters $\{a,\theta,h,M_{\mathrm{BH}}\}$ is similar to the one used by E14. First, we considered three spin values, $a=\{0,0.676,1\}$. For each spin value we considered an ensemble of 18 heights ranging from $2.3$ to $100r_{\mathrm{g}}$. For every such combination we finally considered three values for the viewing angle, $\theta=\{20^{\circ},40^{\circ},60^{\circ}\}$, and 1000 values for $M_{\mathrm{BH}}$ ranging from $0.1\times10^6\,\mathrm{M}_{\odot}$ to $100\times10^6\,\mathrm{M}_{\odot}$ with a step of $0.1\times10^6\,\mathrm{M}_{\odot}$. The parameter space thus consists of a grid of $3\times3\times18\times1000=162,000$ points.

We computed the response functions in the iron line and continuum bands for each point in the parameter space, and used Eq.\,\ref{eq15} to estimate the corresponding model B time-lag spectrum. We then calculated $\chi^2(a,\theta,h,M_{\mathrm{BH}})$ on the parameter space, based on the observed time-lag spectra of each source, according to Eq.\,\ref{eq16}. The resulting grid of $\chi^2$ points was subsequently interpolated quadratically in the parameters $\{a,\theta\}$, and cubically in $\{h,M_{\mathrm{BH}}\}$. We finally used the continuous, interpolated $\chi^2(a,\theta,h,M_{\mathrm{BH}})$ space to obtain $\chi^2_{\mathrm{min}}$, along with the corresponding best-fit values and confidence intervals of the model B parameters.

\subsection{Model A best-fit results} \label{subsec61}

Model A fits the observed time-lag spectra well for all sources. Our best-fit results are listed in Cols.\,2--4 of Table\,\ref{table3}. The best-fit models are shown as continuous brown lines in Figs.\,\ref{fig2} and \ref{fig3}. The observed time-lag spectra of 1H 0707--495, MCG--6-30-15, and Mrk 766 are flat. The fit is thus contrived for these sources, in the sense that the best-fit $s$ value is $\sim0$. In Col.\,3 of Table\,\ref{table3} we therefore list only the upper limit on $s$. We re-fitted the observed time-lag spectra of these sources to a constant delay (i.e. we set $s=0$). The resulting fit is statistically acceptable ($\chi^2_{\mathrm{min}}/\mathrm{dof}=15.9/13$, 19.0/24, and 19.5/14 for 1H 0707--495, MCG--6-30-15, and Mrk 766, respectively), and the best-fit normalization (i.e. the best-fit constant delay in this case) is $A=-29^{+21}_{-20}\,\mathrm{sec}$, $-21\pm9\,\mathrm{sec}$ and $-71\pm19\,\mathrm{sec}$ for 1H 0707--495, MCG--6-30-15, and Mrk 766, respectively. When we assumed a Gaussian distribution for the best-fit $A$ values, the best-fit errors can be used for $s=0$ to estimate the probability of $A=0$ (i.e. the probability that the observed time-lag spectrum is identically zero). We find a probability of 16\%, 2\%, and 0.02\% for 1H 0707--495, MCG--6-30-15, and Mrk 766, respectively (this is a rough estimate and should thus only be considered as indicative).

The time-lag spectra for the remaining sources show evidence of curvature at low frequencies ($\lesssim2\times10^{-4}\,\mathrm{Hz}$), in the sense that model A requires a non-zero best-fit $s$ value.

\subsection{Model B best-fit results} \label{subsec62}

Model B fits the observed time-lag spectra of all sources well. When allowing for all four model B parameters to be free during the fitting procedure, we found that $a$ and $\theta$ are unconstrained, in the sense that even their 68\% confidence interval is larger than the broadest allowed range for the parameter value ($0-1$ and $20^{\circ}-60^{\circ}$ for $a$ and $\theta$, respectively). The reason is the large errors of the observed time-lags and, as discussed in Appendix \ref{appd}, the weak dependence of the model B time-lag spectra on these parameters. Furthermore, for most sources there is a degeneracy between $h$ and $M_{\mathrm{BH}}$, which is caused by the similar dependence of the model B time-lag spectra on these parameters (see Appendix \ref{appd}).

To constrain $a$ and $h$ in the best possible way, we set $\theta=40^{\circ}$ (the mean value found for a similar sample of sources studied by E14) for all sources, and $M_{\mathrm{BH}}$ to the values listed in Col.\,2 of Table\,\ref{table1}. We then repeated the fitting procedure to obtain the best-fit $a$ and $h$ values. The best-fit models are shown as dashed red lines in Figs.\,\ref{fig2} and \ref{fig3}.

Even by fixing $\theta$ and $M_{\mathrm{BH}}$, we found that $a$ cannot be constrained. In the last column of Table\,\ref{table3} we list the $\chi^2_{\mathrm{min}}/\mathrm{dof}$ value when we allowed $a$ to be free during the fitting procedure, while in parentheses we list the corresponding $\chi^2_{\mathrm{min}}$ values when we froze $a$ to the value of 0 and 1, respectively. They are very similar for almost all sources, indicating that we are unable to constrain $a$. MCG--6-30-15 stands as an exception, since for this source we obtained a best-fit $a$ value of $0.3^{+0.3}_{-0.2}$. The upper 95\% level is 0.8, which is somewhat inconsistent with the results obtained by modelling the X-ray spectrum of this source, which requires $a\sim1$ \citep[e.g.][]{2014ApJ...787...83M}. This is due to our choice of $M_{\mathrm{BH}}$ during the fitting procedure. For example, when we set $M_{\mathrm{BH}}=3\times10^6\,\mathrm{M}_{\odot}$ (which is consistent, within the errors, with the value listed in Col.\,2 of Table\,\ref{table1}), the best-fit value of $a$ is 0.4, while the 95\% confidence level ranges from 0 to 1.

Column 5 of Table\,\ref{table3} lists our best-fit results for $h$. The X-ray source height is well defined only for Mrk 766 and Mrk 335. For 1H 0707--495 and MCG--6-30-15 the best-fit $h$ values are $4r_{\mathrm{g}}$ and $2.3r_{\mathrm{g}}$ (which is the lowest allowed fitting value for $h$), respectively. The lower 68\% limit is $2.3r_{\rm g}$ for 1H 0707--495. The upper limit is $20r_{\mathrm{g}}$ and $3r_{\mathrm{g}}$ for 1H 0707--495 and MCG--6-30-15, respectively. For NGC 4051 we obtained a best-fit $h$ value of $17r_{\mathrm{g}}$, with a lower and upper 68\% limit of $2.3r_{\mathrm{g}}$ and $30r_{\mathrm{g}}$, respectively. Given that the lower limit is equal to the lowest value we considered for $h$, we list only the upper limit on $h$ for these three sources.

The best-fit $h$ values for Ark 564 and NGC 7314 are $83r_{\mathrm{g}}$ and $100r_{\mathrm{g}}$ (which is the highest allowed fitting value for $h$), respectively. The X-ray source height is consistent with the value of $100r_{\mathrm{g}}$ for Ark 564. The lower limit on the best-fit $h$ value is $28r_{\mathrm{g}}$ and $82r_{\mathrm{g}}$ for Ark 564 and NGC 7314, respectively. As a result, we list the lower limit of this parameter for these two sources in Table\,\ref{table3}. The high $h$ values arise because the observed time-lag spectra of these sources increase (in magnitude) with decreasing frequency. The lower limit of $h$ for NGC 7314 is higher than for Ark 564 because $M_{\mathrm{BH}}$ in lower in the former source.

However, it is not certain that these two sources have a large X-ray source height. To investigate this further, we fitted their observed time-lag spectra with a model A+B combination. We kept the X-ray source fixed at $h=3.7r_\mathrm{g}$ (the mean height found by E14), set $\theta=40^{\circ}$, fixed $M_{\mathrm{BH}}$ to the respective values listed in Col.\,2 of Table\,\ref{table1}, and let $a=1$. In effect, we kept all the model B parameters fixed to a given value during the fit (as we explained above, we cannot reach a meaningful fit when we let all the model A and B parameters free during the fit) so that the number of degrees of freedom is the same as when we fit the data with model A. Our best-fit results in this case are $A_{\rm Ark\,564}=172^{+62}_{-55}\,\mathrm{sec}$, $A_{\rm NGC\,7314}=69^{+38}_{-41}\,\mathrm{sec}$, $s_{\rm Ark\,564}=1.7^{+0.3}_{-0.4}$, and $s_{\rm NGC\,7314}=1.6^{+1.0}_{-0.6}$. As expected, the reverberation time-lag component causes the resulting best-fit $A$ and $s$ values to be lower and steeper, respectively, than the respective best-fit model A values listed Table\,\ref{table3}. The quality of the combined model A+B fit is similar to that of model A: $\chi^2_{\mathrm{min}}/\mathrm{dof}=14.6/18$, and $\chi^2_{\mathrm{min}}/\mathrm{dof}=26.6/20$ in the case of Ark 564 and NGC 7314, respectively. This result shows that the observed iron line vs continuum time-lags of Ark 564 and NGC 7314 can be fitted well by a combination of a continuum plus reverberation component, the latter of which corresponds to a low $h$ and high $a$ value.

\section{Discussion and conclusions} \label{sec7}

We performed a systematic analysis of the iron line vs continuum ($5-7$ vs $2-4\,\mathrm{keV}$) time-lags in seven AGN. The AGN we studied are X-ray bright and highly variable. The BH mass estimates for these sources are $\lesssim5\times10^6\,\mathrm{M}_{\odot}$, except for Mrk 335, which has a corresponding estimate of $\sim3\times10^7\,\mathrm{M}_{\odot}$ (note that these mass estimates are determined from optical techniques like reverberation mapping, and are not derived here).

Our measurements are among the best that can currently be achieved and are able to be obtained for many years to come (with current X-ray satellites). Our choice of focusing on the iron line band was motivated by the simple fact that its existence indicates the presence of an X-ray reflection component (either from the disc or from distant material) in this band. It is thus is a clean band, ideal for investigating whether X-ray reflection operates in the inner part of the putative accretion disc. However, the low number of photons in this band undermines this advantage. Nevertheless, we found that the iron line vs continuum time-lags are consistent with the simplest X-ray reflection scenario. They also imply X-ray source heights that are close to those derived using data from lower energy bands. This result supports the hypothesis that the X-ray soft excess in these sources is a reflection component (see the relevant discussion in Sect.\,\ref{subsec73}).

\subsection{Estimation of time-lag spectra} \label{subsec71}

We used all the available archival \textit{XMM-Newton} data for seven X-ray bright and highly variable Seyfert galaxies and employed standard Fourier techniques to estimate the iron line vs continuum time-lag spectrum of each source. These sources have a large ($\gtrsim0.3\,\mathrm{Ms}$) amount of archival \textit{XMM-Newton} data. We also took the results obtained from extensive numerical simulations performed by EP16 into account, who studied the effects of the light curve characteristics (duration, time bin size, and Poisson noise) on the statistical properties of the traditional time-lag estimators assuming various intrinsic time-lag spectra commonly observed between X-ray light curves of accreting systems. EP16 found the following: 
\noindent
\begin{itemize}
\item[a)] Time-lag estimates should be computed at frequencies lower than half the Nyquist frequency. This minimises the effects of light-curve binning on their mean values.
\item[b)] The cross-periodogram should not be binned over neighbouring frequencies, as this may introduce significant bias that can only be taken into account when a model CS (and not just a model time-lag spectrum) is assumed. 
\item[c)] Time-lags should be estimated from a cross-periodogram that is averaged over pairs of continuous light-curve segments with the same duration.
\item[d)] If the number of segments, $m$, is $\gtrsim10$, the time-lag estimates will have known errors and approximately follow a Gaussian distribution, provided they are estimated at frequencies at which the sample coherence is $\gtrsim1.2/(1+0.2m)$. This minimises the effects of Poisson noise on their mean values.
\end{itemize}
\noindent
Following these results, we chose the segment duration to be $\sim20\,\mathrm{ks}$. This limits the minimum frequency that can be reliably probed to be $\sim5\times10^{-5}\,\mathrm{Hz}$. A longer segment duration would allow us to probe even lower frequencies, but at the same time it would decrease the number of the available segments, and, consequently, increase the error of the resulting time-lag estimates. According to EP16, if the segment duration is $\gtrsim20\,\mathrm{ks}$, then the time-lag bias should be $\lesssim15\%$ compared to their intrinsic values for the model CS they considered. In Appendix \ref{appe}, we demonstrate that we do not expect the time-lag bias to be a problem in our study.

The maximum frequency that can be reliably probed by the current data is set by point (d) above. The frequency at which the coherence becomes lower than the critical value of $\sim1.2/(1+0.2m)$ depends on the number of segments and is mainly determined by the energy band with the lowest average count rate. This is the iron line band in all cases; the mean count rate of all light curves in our sample is $0.38\pm0.27\,\mathrm{cts/sec}$ and $1.49\pm0.98\,\mathrm{cts/sec}$ for the iron line and continuum band, respectively. We found that the maximum frequency is $\lesssim10^{-3}\,\mathrm{Hz}$ for all sources. Given that the sources in our sample are X-ray bright and have a large amount of archival data, the available \textit{XMM-Newton} data allow for the reliable estimation of iron line vs continuum time-lags at frequencies between $\sim5\times10^{-5}\,\mathrm{Hz}$ and $\sim10^{-3}\,\mathrm{Hz}$.

A direct comparison with published iron line vs continuum time-lags for the sources in our sample is complicated by three factors: the choice of energy bands, the \textit{XMM-Newton} observations used to estimate them, and the cross-periodogram smoothing and/or averaging scheme employed to estimate the time-lags. 

Similar energy bands to ours have been used for Mrk 335, NGC 7314, NGC 4151, and MCG--5-23-16. For Mrk 335, the time-lag magnitudes and errors we find are consisted with those reported by CY15, although they only used data from a single \textit{XMM-Newton} observation, which corresponds to $\sim40\%$ of the data we used. The iron line vs continuum time-lags reported by \citet{2013ApJ...767..121Z} for NGC 7314 are also roughly consistent in magnitude with our findings. They used data from only two \textit{XMM-Newton} observations, which corresponds to $\sim30\%$ of the data we used. Their time-lags are larger (in magnitude) than ours at low frequencies. They provide time-lag estimates at frequencies lower than ours. Owing to the limited length of the data sets they used, their low-frequency estimates must have been obtained from averaging a small number of cross-spectral estimates at neighbouring frequencies. As a result, according to the EP16, these estimates should be far from being Gaussian-distributed, and the frequently used time-lag error prescription of \citet{1999ApJ...510..874N} should severely underestimate the true scatter of these estimates around their mean. We did not estimate time-lags for NGC 4151 and MCG--5-23-16, since the available \textit{XMM-Newton} archival light curves at the time we were analysing the data were not long enough to obtain reliable (in the sense explained in Sect.\,\ref{sec3}) time-lag estimates.

\subsection{Modelling the observed time-lag spectra} \label{subsec72}

We considered two different model time-lag spectra: (a) a power-law time-lag spectrum that describes delays between X-ray continuum variations in different energy bands (model A), and (b) a reverberation time-lag spectrum that describes delays between the X-ray continuum and reprocessed disc emission in a lamp-post geometry (model B). The first is a phenomenological model, while the second is a physical model that depends on the central source geometry. We calculated the model B time-lag spectra by determining accurate disc response functions in the iron line and continuum bands. We fixed the photon index of the X-ray source at a value of 2 and assumed a neutral, prograde disc with an iron abundance equal to the solar value, around a spinning BH. The inner disc radius was set to the location of the ISCO, and the outer radius was fixed at $10^3r_{\mathrm{g}}$. We took all relativistic effects into account and considered the total reprocessed disc emission (and not just the photons initially emitted by the disc at $6.4\,\mathrm{keV}$) in both the iron line and continuum bands. In this respect, our modelling is more accurate than previous attempts (e.g. E14 and C14). 

We found that the model B time-lag spectra have a weak dependence on the BH spin and viewing angle. On the other hand, they depend strongly on the BH mass and X-ray source height. These parameters affect the model B time-lag spectra in a similar way. As the height increases, the model B time-lag spectra flatten at lower frequencies, and to a lower level; the same effect can also be produced by a higher BH mass for the same height (in units of $r_{\mathrm{g}}$). In addition, the characteristic flattening of the reverberation time-lag spectra to a constant value at low frequencies also depends on the outer disc radius. Therefore, the magnitude of this constant level cannot be used in a straight-forward way to determine either the X-ray source height or the outer disc radius, even when the BH mass is known.

Our modelling can be improved in many ways. For example, we could let the slope of the X-ray continuum spectrum, as well as the iron abundance, be free parameters. These parameters mainly influence the amplitude of the disc response function (as they affect the reflection fraction in each energy band). In this case, these parameters should affect the response functions similarly to the BH spin (at small heights). Consequently, we do not expect the difference in the resulting model time-lag spectra to be significant (see the bottom left panel in Fig.\,\ref{figb1}). As shown by CY15, for instance, disc ionisation also affects the model time-lag spectra and should be included in the determination of the response functions. More importantly, however, the main limitation of our modelling is the adopted geometry. The lamp-post geometry is a simplification of the AGN X-ray emitting region. A different geometry can significantly affect the shape and amplitude of the disc response function, and as a result, it can significantly alter the resulting model time-lag spectrum (see the discussion in Appendices \ref{appb} and \ref{appd}). We adopted it (as has been done by many authors in the past) because the estimation of the disc response is relatively straightforward in this case. Furthermore, our intention was to investigate whether the observed iron line vs continuum time-lag spectra are consistent with the simplest theoretical reverberation model, and to see which constraints they can impose on the X-ray source and disc geometry. In retrospect, given the results of our study (see the discussion below), the current data sets fail to distinguish between the predictions of the lamp-post model and those from a more detailed approach.

\subsection{Model-fit results} \label{subsec73}

We fitted models A and B separately to the observed time-lag spectra because given their quality (limited frequency range and large errors), we would not have been able to constrain the lamp-post parameters by fitting a combined model A+B to the data. Both models provide statistically acceptable fits. We therefore cannot prefer one model based on the quality of the model fits.

However, our best-fit results do provide useful hints. For example, the best-fit model A power-law index values for 1H 0707--495, MCG--6-30-15, and Mrk 766 are consistent with zero. The observed time-lags in these sources are flat, and the best-fit model A reduces to just a constant. This result (i.e. that the best-fit power-law model to the data is a horizontal line) leads us to believe that the case for X-ray reverberation time-lags is strong, at least in these three sources. If the observed time-lags were indeed representative of continuum time-lags, we would expect a non-zero best-fit slope.

As we showed in Sect.\,\ref{subsec43}, the observed time-lags should have both a continuum and a reverberation component. The lack of a significant detection of the expected continuum component for these three sources (at least) is not surprising and can be explained physically. The continuum time-lags depend on the energy separation between the chosen energy bands, which is  small in our case. Our best-fit model A amplitude values are systematically lower than the respective best-fit values found by E14. This is what we should expect for continuum time-lags, as the energy separation between the iron line and continuum bands is smaller than the separation between the $1.5-4$ and $0.3-1\,\mathrm{keV}$ bands used by E14.

When fitting the observed time-lags to the model B time-lag spectrum, we found that the BH spin and inclination cannot be constrained. This is due to the large errors of the time-lag estimates and the weak dependence of the model B time-lag spectra on these parameters. Furthermore, there is a degeneracy between the X-ray source height and the BH mass that is due to the similar dependence of the model B time-lag spectrum on these parameters. We thus froze the BH mass value for each source to the most accurate and reliable values we could find in the literature and managed to constrain the X-ray source height. The observed iron line vs continuum time-lag spectra either require, or are consistent with, small X-ray source heights. For example, the best-fit height estimates are $\lesssim10r_g$ in three sources. The best-fit height for NGC 4051 is also consistent with such a low value. Even for Ark 564 and NGC 7314, the data are consistent with an X-ray source height as small as $\sim4r_{\mathrm{g}}$ when we considered a combined model A+B.

\begin{figure}[h!]
   \centering
   \includegraphics[width=\hsize]{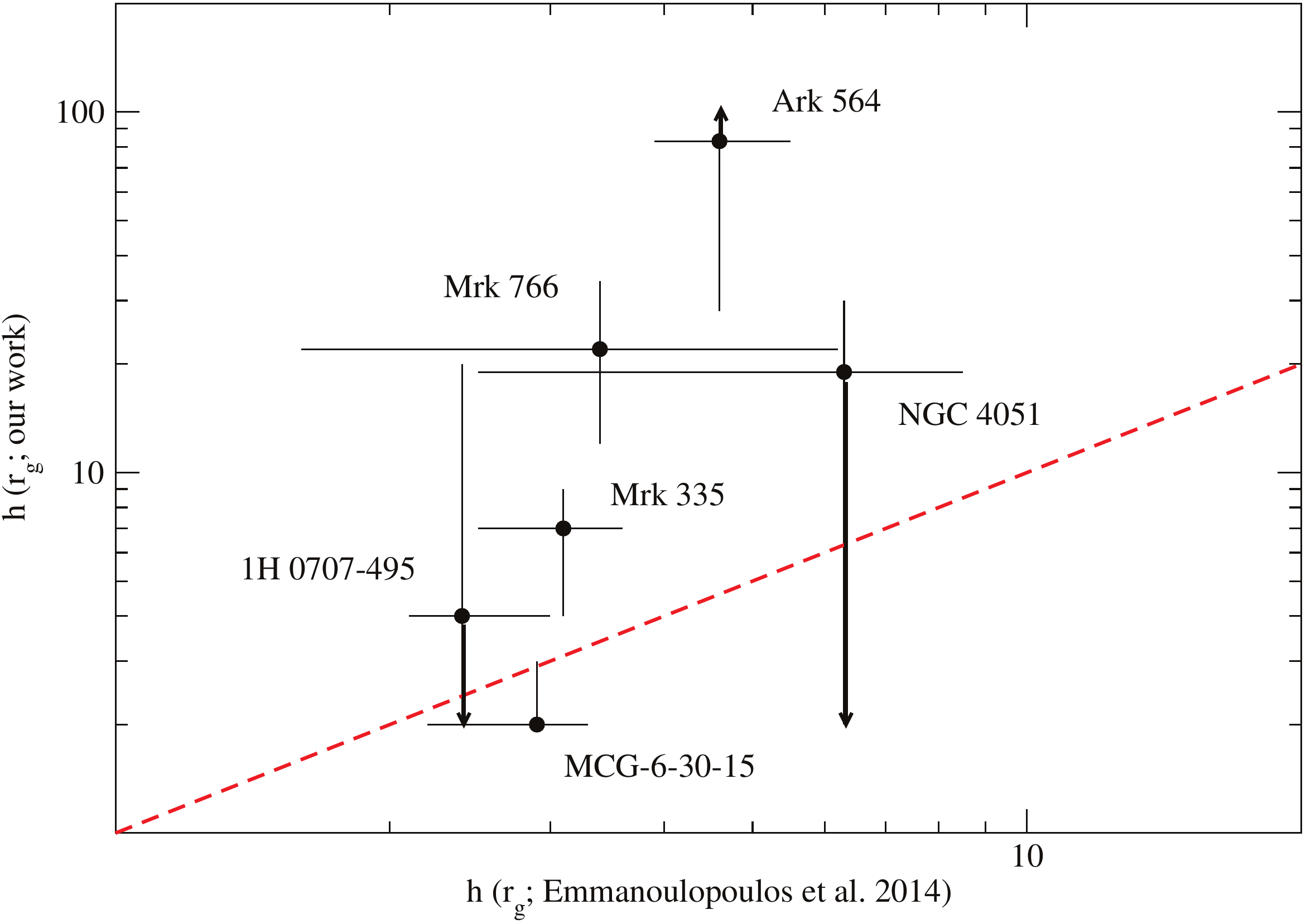}
      \caption{Comparison between the best-fit X-ray source height obtained by fitting the iron line vs continuum time-lags (vertical axis; this work) with those obtained by fitting the soft excess vs continuum time-lags (horizontal axis; E14).}
         \label{fig4}
   \end{figure}

Figure\,\ref{fig4} shows the our best-fit $h$ values versus the E14 best-fit results. The red dashed line indicates the one-to-one relation. Although most of the points are located above this line, given the large uncertainties, the plot suggests a broad agreement with the results of E14. The direct comparison is complicated because we considered more data sets than E14 for some sources. \citet{2013MNRAS.435.1511A} showed that the soft lags of NGC 4051 vary significantly and systematically with source flux. In our case, we cannot fit model B to time-lag spectra estimated from low- and high-flux segments, as the uncertainty on the model parameters will be significantly larger than what we obtain when we fit the overall time-lags. Nevertheless, if this trend is present in all AGN and in the iron line vs continuum time-lags as well, then when we average over data with a wide flux range, segments with the highest flux may dominate the cross-periodogram, as they may be associated with higher amplitude variations \citep[due to the rms-flux relation;][]{2001MNRAS.323L..26U}. If the data sets we considered exhibit a wider flux variability range than the one in the E14 data sets, differences in the best-fit results may be easier to explain.
 
In conclusion, the soft excess vs continuum time-lags are consistent with the iron line vs continuum time-lags we presented here, in that they both support the hypothesis of disc reflection from an X-ray source that is located very close to the disc and the central BH.

\subsection{Implications for the X-ray reflection scenario} \label{subsec74}

Except for the source height, we are unable to constrain additional reverberation model parameters such as the BH mass and spin, viewing angle, and the outer disc radius. Accurate determination of these parameters would require a significant reduction in the errors of the time-lag estimates and/or an increase in the frequency range that can be reliable probed. However, this requires a substantial increase in the number of X-ray observations of AGN.

For example, to probe frequencies lower by a factor of $\sim5$ (i.e. to reach a low limit of $\sim10^{-5}\,\mathrm{Hz}$), segments with a duration of $\sim100\,\mathrm{ks}$ are required. Assuming the number of segments used for the time-lag estimation remains the same as in the present work, this would require the net \textit{XMM-Newton} exposure times to increase by a factor of $\sim5$ for each source (on average). This will, however, neither decrease the error of the time-lag estimates nor allow allow us to probe higher frequencies, since both require an increase in the number of segments. Extending the high-frequency limit requires an increase of $\nu_{\mathrm{max}}$, which can only be achieved by increasing the number of segments. For example, to probe frequencies $\sim2\times10^3\,\mathrm{Hz}$ for MCG--6-30-15, the critical coherence value has to decrease from its present value of $\sim0.18$ to $\sim0.05$ (see Fig.\,\ref{fig1}). This requires the number of segments to increase from 28 to 115, which corresponds to an increase in the net \textit{XMM-Newton} exposure times by a factor of $\sim4$. This would, in turn, reduce the errors of the time-lag estimates by a factor of $\sim2$. In this case, however, we would be unable to probe lower frequencies, since this requires segments of longer duration.

One possibility to extend the frequency range of the observed time-lag spectra would be to use the large volume of available archival data from past and current low-Earth orbit satellites (e.g. \textit{ASCA}, \textit{Chandra}, and \textit{Suzaku}). The idea would be to bin the respective light curves at one orbital period ($\sim96\,\mathrm{min}$) to probe low frequencies, although this requires a large number of long observations. For instance, estimating time-lags at frequencies lower than $\sim10^{-5}\,\mathrm{Hz}$ requires an ensemble of at least ten observations, which will be longer than at least a few days. We are currently investigating this possibility to estimate time-lag spectra over a wider frequency range.

Given the quality of the present data sets in the iron line band and the resulting iron line vs continuum time-lag spectra, the need for constructing more sophisticated theoretical disc response functions is questionable. It seems that the best way to test the X-ray reverberation scenario and significantly constrain the model parameters is to focus on the soft excess vs continuum time-lag modelling, where the S/N of the existing light curves in the soft band is much higher than those in the iron line band. This would require considering the ionisation structure of the disc in the construction of appropriate disc response functions.

Modelling the energy dependence of the time-lag spectra is another possibility. However, we note that the errors of the resulting time-lag estimates are dictated by the energy band with the lower average count rate. As such, the use of light curves over a broad energy band as a reference should not significantly lower the errors of the resulting time-lag estimates, even at the lowest possible frequencies. We plan to model the energy dependence of the observed time-lag spectra in a future work, where we will also consider \textit{NuSTAR} data to study time-lags between the Compton hump and the X-ray continuum.

\begin{acknowledgements} 
We thank the referee for his/her suggestions, which significantly improved the quality and clarity of the manuscript. This work was supported  by the AGNQUEST project, which is implemented under the Aristeia II Action of the Education and Lifelong Learning operational programme of the GSRT, Greece. The research leading to these results has also received funding from the European Union Seventh Framework Programme (FP7/2007-2013) under grant agreement n$^{\rm o}$ 312789, and by the grant PIRSES-GA-2012-31578 EuroCal.
\end{acknowledgements} 

\bibliographystyle{aa}
\bibliography{refs}

\begin{appendix}

\section{Model CS} \label{appa}

Substituting Eqs.\,\ref{eq10} and \ref{eq11} into Eq.\,\ref{eq12}, we can compute the CCF between the iron line and continuum bands as follows:
\noindent
\begin{align} \label{eqa1}
\nonumber
R_{5-7,2-4}(\tau)= & R^{(\mathrm{c})}_{5-7,2-4}(\tau) \\
\nonumber
& +\int_{-\infty}^{\infty}\Psi_{2-4}(t')R^{(\mathrm{c})}_{5-7,2-4}(\tau+t')\mathrm{d}t' \\
\nonumber
& +\int_{-\infty}^{\infty}\Psi_{5-7}(t')R^{(\mathrm{c})}_{5-7,2-4}(\tau-t')\mathrm{d}t' \\
\nonumber
& +\int_{-\infty}^{\infty}\int_{-\infty}^{\infty}\Psi_{2-4}(t')\Psi_{5-7}(t'') \\
& \quad\times R^{(\mathrm{c})}_{5-7,2-4}(\tau+t'-t'')\mathrm{d}t'\mathrm{d}t'',
\end{align}
\noindent
where $R^{(\mathrm{c})}_{5-7,2-4}(\tau)\equiv\mathrm{E}\{[\mathscr{F}^{(\mathrm{c})}_{2-4}(t)-\mu_{2-4}][\mathscr{F}^{(\mathrm{c})}_{5-7}(t+\tau)-\mu_{5-7}]\}$ is the CCF of the continuum emission. Consequently, the intrinsic CS is
\noindent
\begin{align} \label{eqa2}
\nonumber
C_{5-7,2-4}(\nu) &\equiv\int_{-\infty}^{\infty}R_{5-7,2-4}(\tau)\mathrm{e}^{-\mathrm{i}2\pi\nu\tau}\mathrm{d}\tau \\
\nonumber
&=\int_{-\infty}^{\infty}R^{(\mathrm{c})}_{5-7,2-4}(\tau)\mathrm{e}^{-\mathrm{i}2\pi\nu\tau}\mathrm{d}\tau \\
\nonumber
& \quad +\int_{-\infty}^{\infty}\int_{-\infty}^{\infty}\Psi_{2-4}(t')R^{(\mathrm{c})}_{5-7,2-4}(\tau+t')\mathrm{e}^{-\mathrm{i}2\pi\nu\tau}\mathrm{d}t'\mathrm{d}\tau \\
\nonumber
& \quad +\int_{-\infty}^{\infty}\int_{-\infty}^{\infty}\Psi_{5-7}(t')R^{(\mathrm{c})}_{5-7,2-4}(\tau-t')\mathrm{e}^{-\mathrm{i}2\pi\nu\tau}\mathrm{d}t'\mathrm{d}\tau \\
\nonumber
& \quad +\int_{-\infty}^{\infty}\int_{-\infty}^{\infty}\int_{-\infty}^{\infty}\Psi_{2-4}(t')\Psi_{5-7}(t'') \\
& \quad\quad \times R^{(\mathrm{c})}_{5-7,2-4}(\tau+t'-t'')\mathrm{e}^{-\mathrm{i}2\pi\nu\tau}\mathrm{d}t'\mathrm{d}t''\mathrm{d}\tau.
\end{align}
\noindent
Setting $C^{(\mathrm{c})}_{5-7,2-4}(\nu)\equiv\int_{-\infty}^{\infty}R^{(\mathrm{c})}_{5-7,2-4}(\tau)\mathrm{e}^{-\mathrm{i}2\pi\nu\tau}\mathrm{d}\tau$ and applying the convolution theorem, Eq.\,\ref{eqa2} becomes
\begin{align} \label{eqa3}
\nonumber
C_{5-7,2-4}(\nu) &=C^{(\mathrm{c})}_{5-7,2-4}(\nu) \\
\nonumber
& \quad +\tilde{\Psi}^{*}_{2-4}(\nu)C^{(\mathrm{c})}_{5-7,2-4}(\nu) \\
\nonumber
& \quad +\tilde{\Psi}_{5-7}(\nu)C^{(\mathrm{c})}_{5-7,2-4}(\nu) \\
\nonumber
& \quad +\tilde{\Psi}^{*}_{2-4}(\nu)\tilde{\Psi}_{5-7}(\nu)C^{(\mathrm{c})}_{5-7,2-4}(\nu) \\
&=C^{(\mathrm{c})}_{5-7,2-4}(\nu)[1+\tilde{\Psi}_{5-7}(\nu)][1+\tilde{\Psi}_{2-4}(\nu)]^{*}.
\end{align}
\noindent
Subsequently, the model time-lag spectrum is given by
\noindent
\begin{align} \label{eqa4}
\nonumber
\tau_{5-7,2-4}(\nu) &\equiv\frac{1}{2\pi\nu}\mathrm{arg}\{C^{(\mathrm{c})}_{5-7,2-4}(\nu)[1+\tilde{\Psi}_{5-7}(\nu)][1+\tilde{\Psi}_{2-4}(\nu)]^{*}\} \\
\nonumber
&=\frac{\mathrm{arg}[C^{(\mathrm{c})}_{5-7,2-4}(\nu)]+\mathrm{arg}\{[1+\tilde{\Psi}_{5-7}(\nu)][1+\tilde{\Psi}_{2-4}(\nu)]^{*}\}}{2\pi\nu} \\
&=\tau_{5-7,2-4}^{(\mathrm{c})}(\nu)+\tau_{5-7,2-4}^{(\mathrm{r})}(\nu),
\end{align}
\noindent
where we used the property $\mathrm{arg}[z_1z_2]=\mathrm{arg}[z_1]+\mathrm{arg}[z_2]$, for the complex numbers $z_1$, $z_2$. The functions appearing on the right-hand side of Eq.\,\ref{eqa4} are defined as $\tau_{5-7,2-4}^{(\mathrm{c})}(\nu)\equiv(2\pi\nu)^{-1}\mathrm{arg}[C^{(\mathrm{c})}_{5-7,2-4}(\nu)]$, and $\tau_{5-7,2-4}^{(\mathrm{r})}(\nu)\equiv(2\pi\nu)^{-1}\mathrm{arg}\{[1+\tilde{\Psi}_{5-7}(\nu)][1+\tilde{\Psi}_{2-4}(\nu)]^{*}\}$. The total time-lag spectrum is therefore equal to the sum of two time-lag spectra; one due to delays between variations of different energy bands in the continuum, $\tau_{5-7,2-4}^{(\mathrm{c})}(\nu)$, and one due to delays between the reprocessed disc emission and the continuum, $\tau_{5-7,2-4}^{(\mathrm{r})}(\nu)$.
The function $\tau_{5-7,2-4}^{(\mathrm{r})}(\nu)$ can be written as follows:
\noindent
\begin{align} \label{eqa5}
\nonumber
\tau^{(\mathrm{r})}_{5-7,2-4}(\nu) &\equiv\frac{1}{2\pi\nu}\mathrm{arg}\{[1+\tilde{\Psi}_{5-7}(\nu)][1+\tilde{\Psi}_{2-4}(\nu)]^{*}\} \\
&=\frac{1}{2\pi\nu}\mathrm{arctan}\left\{\frac{\Im\{[1+\tilde{\Psi}_{5-7}(\nu)][1+\tilde{\Psi}^{*}_{2-4}(\nu)]\}}{\Re\{[1+\tilde{\Psi}_{5-7}(\nu)][1+\tilde{\Psi}^{*}_{2-4}(\nu)]\}}\right\},
\end{align}
\noindent
where
\noindent
\begin{align} \label{eqa6}
\nonumber
\Re\{[1+\tilde{\Psi}_{5-7}(\nu)][1+\tilde{\Psi}^{*}_{2-4}(\nu)]\}= & 1+\Re[\tilde{\Psi}_{2-4}(\nu)]+\Re[\tilde{\Psi}_{5-7}(\nu)] \\
\nonumber
& +\Re[\tilde{\Psi}_{2-4}(\nu)]\Re[\tilde{\Psi}_{5-7}(\nu)] \\
& +\Im[\tilde{\Psi}_{2-4}(\nu)]\Im[\tilde{\Psi}_{5-7}(\nu)],
\end{align}
\noindent
\begin{align} \label{eqa7}
\nonumber
\Im\{[1+\tilde{\Psi}_{5-7}(\nu)][1+\tilde{\Psi}^{*}_{2-4}(\nu)]\}= & \Im[\tilde{\Psi}_{2-4}(\nu)]-\Im[\tilde{\Psi}_{5-7}(\nu)] \\
\nonumber
& -\Re[\tilde{\Psi}_{2-4}(\nu)]\Im[\tilde{\Psi}_{5-7}(\nu)] \\
& +\Im[\tilde{\Psi}_{2-4}(\nu)]\Re[\tilde{\Psi}_{5-7}(\nu)],
\end{align}
\noindent
$\Re[\tilde{\Psi}_{\mathcal{E}}(\nu)]=\int_{-\infty}^{\infty}\Psi_{\mathcal{E}}(t)\cos(2\pi\nu t)\mathrm{d}t$, and $\Im[\tilde{\Psi}_{\mathcal{E}}(\nu)]=-\int_{-\infty}^{\infty}\Psi_{\mathcal{E}}(t)\sin(2\pi\nu t)\mathrm{d}t$. In other words, the model time-lag spectrum depends in a non-trivial way on the Fourier transforms of the iron line and continuum response functions.

\section{Disc response functions} \label{appb}

\begin{figure*}[ht!]
\centering
\includegraphics[width=15.0cm,clip]{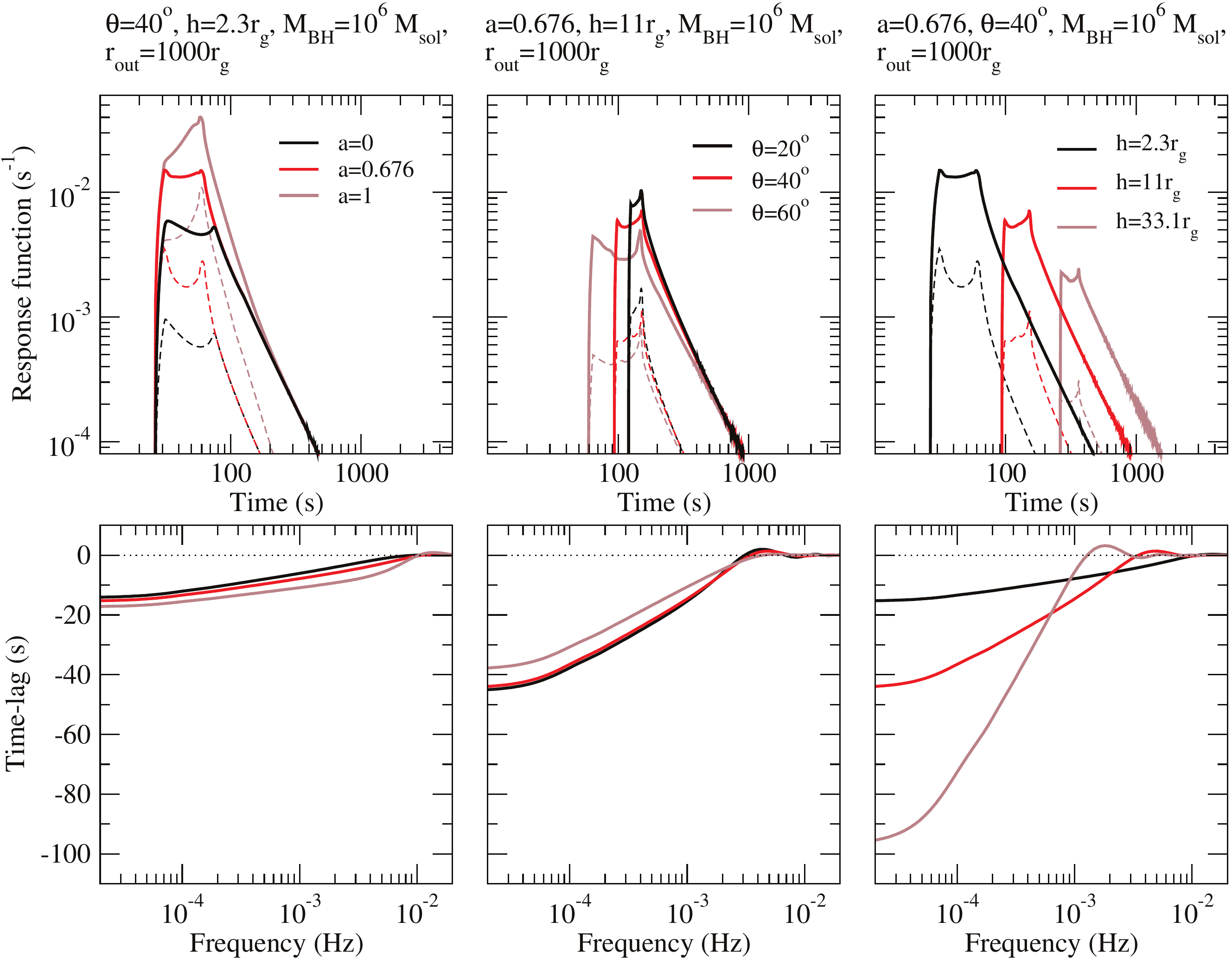}
\caption{Dependence of the $5-7\,\mathrm{keV}$ (top panels; continuous lines) and $2-4\,\mathrm{keV}$ (top panels; dashed lines) disc response functions on $a$ (top left panel), $\theta$ (top middle panel) and $h$ (top right panel), along with the corresponding time-lag spectra (bottom panels). The response functions have been normalised such that their area is equal to the observed ratio of reprocessed-to-continuum photons.}
\label{figb1}
\end{figure*}

The top panels in Fig.\,\ref{figb1} show various disc response functions in the energy bands $5-7\,\mathrm{keV}$, $\Psi_{5-7}(t)$ (solid lines), and $2-4\,\mathrm{keV}$, $\Psi_{2-4}(t)$ (dashed lines), in the case when the disc extends to $10^3r_{\mathrm{g}}$, and $M_{\rm BH}=10^6\,\mathrm{M}_{\odot}$. The horizontal axes show time in the observer frame, with the origin ($t=0$) corresponding to the beginning of the primary X-ray flare (which, as we mentioned in Sect.\,\ref{subsec44}, ends abruptly at $t=1t_{\mathrm{g}}$). The response functions plotted in these panels are defined such that $\int_{-\infty}^{\infty}\Psi_{\mathcal{E}}(t)\mathrm{d}t$ is equal to the observed ratio of reprocessed-to-continuum photons. 

The parameters $a$, $\theta$, $h$, and $r_{\mathrm{out}}$ affect both the shape and amplitude of the response function. The parameter $M_{\mathrm{BH}}$ changes its shape (by either uniformly stretching or contracting it in the horizontal direction, depending on whether $M_{\mathrm{BH}}$ is increased or decreased), but not their amplitude. In all cases the response function shows an initial sharp rise at a time $t_{\mathrm{rise}}>0$, followed by a second peak at a later times, and a final gradual decline up to a maximum time $t_{\mathrm{max}}$. The initial rise time corresponds to the instant the observer detects the first reflected emission from the near side of the disc. The second peak appears when the observer detects emission from the far side of the disc. At longer timescales we detect emission from the outer disc radii, where the reflection amplitude is reduced, and hence $\Psi_{\mathcal{E}}(t)$ gradually decreases up until $t_{\mathrm{max}}$, when we detect the last emission from the edge of the far side of the disc, and $\Psi_{\mathcal{E}}(t)$ abruptly drops to zero. $t_{\mathrm{max}}$ depends mainly on $h$ and $r_{\mathrm{out}}$. If $r_{\mathrm{out}}\rightarrow\infty$, then $t_{\mathrm{max}}\rightarrow\infty$ and the response function has a $\sim t^{-2}$ behaviour at long times \citep[e.g.][]{2013MNRAS.430..247W}.

Given our adopted convention, the amplitude of the response function in a given energy band depends on the strength of the reprocessed disc emission relative to the continuum emission in that energy band. Therefore, in almost all cases shown in Fig.\,\ref{figb1}, the amplitude of $\Psi_{5-7}(t)$ is larger than the amplitude of $\Psi_{2-4}(t)$ by a factor of $\sim10$. However, for high spin values and small heights, this difference is as small as $\sim4$ (top left panel in Fig.\,\ref{figb1}). Since the amplitude of both $\Psi_{5-7}(t)$ and $\Psi_{2-4}(t)$ affect the model time-lag spectra (see Appendix \ref{appa}), $\Psi_{2-4}(t)$ should not be neglected when estimating the theoretical time-lag spectra.

The BH spin affects neither the width of the response function, nor $t_{\mathrm{rise}}$ (top left panel in Fig.\,\ref{figb1}). This may seem counter-intuitive at first, as $r_{\mathrm{ISCO}}$ decreases from $6r_{\mathrm{g}}$ for a non-spinning BH, to almost $1r_{\mathrm{g}}$ for a maximally spinning BH, reducing the distance between the X-ray source and the inner disc. However, for the $\theta$ and $h$ values we considered, the reprocessed emission that is initially detected by the distant observer is emitted from a part of the disc that is located towards the observer's direction at a radius larger than $6r_{\mathrm{g}}$, independent of $a$. Nevertheless, $a$ does affect the amplitude of the response function, especially when $h$ is small. When the X-ray source is located close to the BH, light-bending effects are strong, and most of the continuum emission illuminates the region of the disc close to the BH. For a maximally spinning BH the disc extends to very small radii, and hence the amplitude of the response function when $a=1$ is significantly larger than when $a=0$ (brown and black lines, respectively, in the top left panel of Fig.\,\ref{figb1}).

The rise time decreases and the width of the response function increases with increasing inclination (top middle panel in Fig.\,\ref{figb1}). The rise time decrease is due to the decreased light path difference between the continuum and the disc emission with increasing $\theta$. The increase in the width appears because the difference between the light travel time from the near-side of the disc and from the X-ray source increases with increasing $\theta$. The amplitude of the response function decreases with increasing $\theta$, since the projected area of the disc (and hence the observed amount of reflected emission) is proportional to $\cos\theta$.

The top right panel in Fig.\,\ref{figb1} shows that $t_{\mathrm{rise}}$ depends mainly on the height of the X-ray source, as it strongly increases with increasing $h$. The reason is the increase in the light travel time between the X-ray source and the disc emission. The width of the response function also increases with increasing $h$. As $h$ increases, the difference between $t_{\mathrm{rise}}$ and the time when the second peak appears increases as well because the light travel time difference between the near and far-side of the disc increases.

\section{Effects of finite continuum flare duration on the response function estimation} \label{appc}

As mentioned in Sect.\,\ref{subsec44}, the disc response functions were determined assuming a flare of constant flux continuum emission that lasted for $1t_{\mathrm{g}}$. Our response functions thus formally describe the response of the disc to a flare with a finite duration and a box-like light curve (instead of a delta-function), and we henceforth designate them as $\Psi^{\mathrm{(b)}}_{\mathcal{E}}(t)$. In this appendix we investigate the relation between the Fourier transforms of $\Psi_{\mathcal{E}}(t)$ (i.e. the disc response to an instantaneous continuum flare) and $\Psi^{(\mathrm{b})}_{\mathcal{E}}(t)$. This is necessary, because the model time-lag spectra given by Eq.\,\ref{eq15} depend on the Fourier transforms of $\Psi_{\mathcal{E}}(t)$ in the iron line and continuum bands.

We assumed that $\mathscr{F}^{(\mathrm{c})}_{\mathcal{E}}(t)$, as it appears on the right-hand side of Eq.\,\ref{eq10}, is equal to $F_0[H(t)-H(t-t_{\mathrm{g}})]t^{-1}_{\mathrm{g}}$, where $H(x)$ is the Heaviside step function ($H(x)$ is defined as being equal to unity when $x\ge0$, and zero otherwise) and $F_0$ is a constant. According to Eq.\,\ref{eq10}, the observed flux is then
\noindent
\begin{align} \label{eqc1}
\nonumber
\mathscr{F}_{\mathcal{E}}(t) &=F_0[H(t)-H(t-t_{\mathrm{g}})]t^{-1}_{\mathrm{g}}+F_0t^{-1}_{\mathrm{g}}\int_{t}^{t+t_{\mathrm{g}}}\Psi_{\mathcal{E}}(t')\mathrm{d}t' \\
& =F_0[H(t)-H(t-t_{\mathrm{g}})]t^{-1}_{\mathrm{g}}+F_0\Psi^{(\mathrm{b})}_{\mathcal{E}}(t),
\end{align}
\noindent
where $\Psi^{(\mathrm{b})}_{\mathcal{E}}(t)=t_{\mathrm{g}}^{-1}\int_{t}^{t+t_{\mathrm{g}}}\Psi_{\mathcal{E}}(t')\mathrm{d}t'$. When $t_{\mathrm{g}}\rightarrow0$ (i.e. when the flare becomes instantaneous), $\Psi^{(\mathrm{b})}_{\mathcal{E}}(t)\rightarrow\Psi_{\mathcal{E}}(t)$, as expected.

According to Eq.\,\ref{eqc1}, $\Psi^{(\mathrm{b})}_{\mathcal{E}}(t)$ is equal to the convolution of $\Psi_{\mathcal{E}}(t)$ with a constant kernel that is non-zero for $0\le t\le t_{\mathrm{g}}$. The relation between the Fourier transforms of the two functions is therefore
\noindent
\begin{align} \label{eqc2}
\nonumber
\tilde{\Psi}_{\mathcal{E}}^{(\mathrm{b})}(\nu) &=\int_{-\infty}^{\infty}\left[\frac{1}{t_{\mathrm{g}}}\int_{t}^{t+t_{\mathrm{g}}}\Psi_{\mathcal{E}}(t')\mathrm{d}t'\right]\mathrm{e}^{-\mathrm{i}2\pi\nu t}\mathrm{d}t \\
& =[\mathrm{e}^{\mathrm{i}\pi\nu t_{\mathrm{g}}}\mathrm{sinc}(\pi\nu t_{\mathrm{g}})]\tilde{\Psi}_{\mathcal{E}}(\nu).
\end{align}
\noindent
In other words, the Fourier transform of $\Psi^{(\mathrm{b})}_{\mathcal{E}}(t)$ has to be divided by the factor $\mathrm{e}^{\mathrm{i}\pi\nu t_{\mathrm{g}}}\mathrm{sinc}(\pi\nu t_{\mathrm{g}})$ to account for the finite width of the continuum flare. This correction term becomes important only at frequencies $\nu\gtrsim t^{-1}_{\mathrm{g}}$, which corresponds to $\gtrsim0.2\,\mathrm{Hz}$ for $M_{\mathrm{BH}}=10^6\,\mathrm{M}_{\odot}$. As explained in Sect.\,\ref{sec2}, time-lags cannot be reliably estimated at such high frequencies with current data, hence the correction term has a negligible effect on the model time-lag spectra in our work. Nevertheless, we applied Eq.\,\ref{eqc2} to determine $\tilde{\Psi}_{5-7}(\nu)$ and $\tilde{\Psi}_{2-4}(\nu)$, which were subsequently used to calculate the corresponding model iron line vs continuum time-lag spectrum according to Eq.\,\ref{eq15}.

\section{Lamp-post model time-lag spectra} \label{appd}

\begin{figure}[h!]
   \centering
   \includegraphics[width=\hsize]{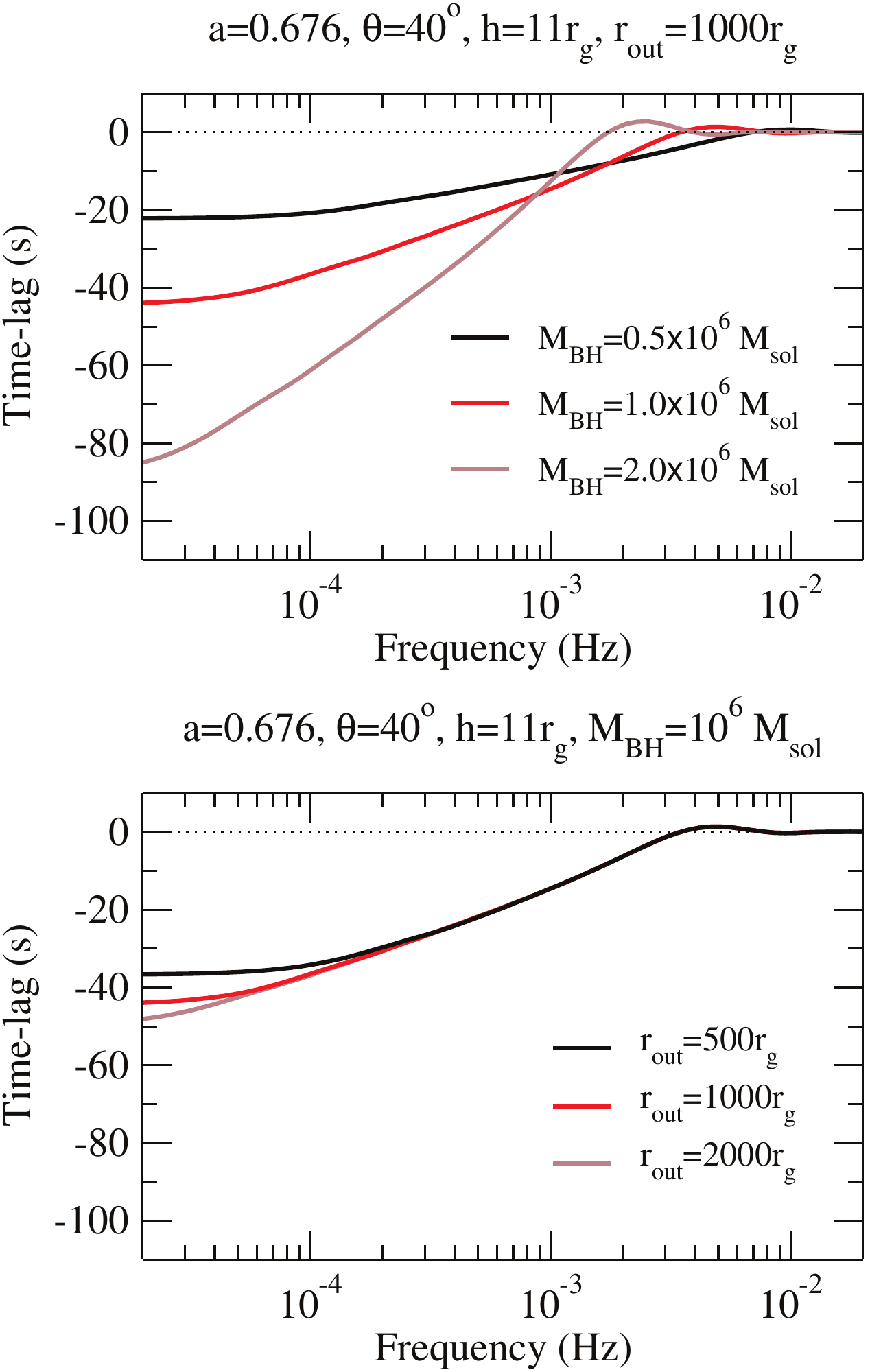}
      \caption{\textit{Top panel}: Model time-lag spectra between the $5-7$ and $2-4\,\mathrm{keV}$ bands for various $M_{\mathrm{BH}}$ values. The remaining model parameters are fixed at $a=0.676$, $\theta=40^{\circ}$, $h=11r_{\mathrm{g}}$, and $r_{\mathrm{out}}=10^3r_{\mathrm{g}}$. \textit{Bottom panel}: Model time-lag spectra between the $5-7\,\mathrm{keV}$ and $2-4\,\mathrm{keV}$ bands for various values of $r_{\mathrm{out}}$. The remaining model parameters are fixed at $a=0.676$, $\theta=40^{\circ}$, $h=11r_{\mathrm{g}}$, and $M_{\mathrm{BH}}=10^6\,\mathrm{M}_{\odot}$.}
         \label{figd1}
   \end{figure}

The bottom panels in Fig.\,\ref{figb1} (and both panels in Fig.\,\ref{figd1}) show various iron line vs continuum model time-lag spectra, $\tau^{(\mathrm{r})}_{5-7,2-4}(\nu)$, calculated using the parameters that we used to estimate the response functions appearing in the same figure. The time-lags  are predominantly negative, meaning that variations in the iron line band are delayed with respect to variations in the continuum band. They all share similar characteristics. They all flatten to a constant, negative plateau, $\tau_{\mathrm{plateau}}$, below a frequency $\nu_{\mathrm{plateau}}$. At higher frequencies they rise to a maximum positive bump, $\tau_{\mathrm{bump}}$, at a frequency $\nu_{\mathrm{bump}}$, followed by sinusoidal behaviour with decreasing amplitude around a zero time-lag value.

As in the case of the response functions, the parameters $a$, $\theta$, $h$, $M_{\mathrm{BH}}$, and $r_{\mathrm{out}}$ affect both the shape and magnitude of the time-lags. The bottom left panel in Fig.\,\ref{figb1} shows that the BH spin has a weak effect on the model time-lag spectra. The time-lags increase slightly in (absolute) magnitude with increasing $a$. The frequency $\nu_{\mathrm{plateau}}$ does not depend on $a$, while $\nu_{\mathrm{bump}}$ and $\tau_{\mathrm{bump}}$ depend weakly on $a$. The dependence of the model time-lag spectra on $a$ is in contrast to the strong dependence of the response function amplitude on the same parameter (top left panel in Fig.\,\ref{figb1}). This result shows that the response function amplitude does not significantly influence the time-lag characteristics. It also highlights the importance of including the reprocessed emission in the continuum band as well (as we did here), since response functions of comparable magnitudes in two bands will lead to a model time-lag spectrum that is close to zero.

In general, when $\Psi_{5-7}(t)=\Psi_{2-4}(t)\ne0$ (i.e. when the observed ratio of reprocessed-to-continuum photons is equal in the two energy bands), Eq.\,\ref{eqa5} reduces to $\tau^{(\mathrm{r})}_{5-7,2-4}(\nu)=0$. As we mentioned in Appendix \ref{appb}, although $\Psi_{5-7}(t)$ increases with increasing $a$, so does the amplitude of $\Psi_{2-4}(t)$, and in fact even more so. For a rapidly spinning BH and X-ray reflection from the innermost part of the disc, the red wing of the iron line is well into the $2-4\,\mathrm{keV}$ band due to strong gravitational effects. Including reprocessed emission in the continuum band has the well-known effect of decreasing, or diluting, the magnitude of the model time-lag spectrum \citep[e.g.][]{2013MNRAS.430..247W}. This can be seen by taking the limit $\Psi_{2-4}(t)\rightarrow0$ in Eqs.\,\ref{eqa6} and \ref{eqa7}, in which case $|\Re\{[1+\tilde{\Psi}_{5-7}(\nu)][1+\tilde{\Psi}^{*}_{2-4}(\nu)]\}|$ is minimised, $|\Im\{[1+\tilde{\Psi}_{5-7}(\nu)][1+\tilde{\Psi}^{*}_{2-4}(\nu)]\}|$ is maximised, and hence the $|\tau^{(\mathrm{r})}_{5-7,2-4}(\nu)|$ is maximised according to Eq.\,\ref{eqa5}.

The bottom middle panel in Fig.\,\ref{figb1} shows that the time-lags increase in magnitude with decreasing inclination. The reason is that the rise time of the response functions decreases as $\theta$ increases. Similarly to the BH spin, $\nu_{\mathrm{plateau}}$ and $\nu_{\mathrm{bump}}$ have a weak dependence on $\theta$.

The (absolute) magnitude of $\tau_{\mathrm{plateau}}$ in the middle bottom panel of Fig.\,\ref{figb1} is larger than the corresponding magnitude in the bottom left panel because we considered a larger X-ray source height in the former case. In a lamp-post geometry, the time-lag spectra are mainly affected by X-ray source height (bottom right panel in the same figure). The (absolute) magnitude of the time-lags and $\tau_{\mathrm{plateau}}$ strongly increase and $\nu_{\mathrm{plateau}}$ and $\nu_{\mathrm{bump}}$ decrease with increasing $h$. Both effects are caused mainly by the fact that $t_{\rm{rise}}$ increases substantially with increasing source height. The (absolute) magnitude of $\tau_{\mathrm{bump}}$ also increases with increasing $h$, although this increase is not as pronounced as in the case of $\tau_{\mathrm{plateau}}$.

The BH mass likewise strongly affects the magnitude and shape of the time-lag spectrum, as seen in the top panel of Fig.\,\ref{figd1}. At a given X-ray source height, the magnitude of the time-lags increases while $\nu_{\mathrm{plateau}}$ and $\nu_{\mathrm{bump}}$ decrease with increasing $M_{\mathrm{BH}}$. This is due to the increased light travel time between the X-ray source and the disc, since the physical size of the X-ray source/disc system scales proportionally with $M_{\mathrm{BH}}$. In other words, since time-lags scale with $t_{\mathrm{g}}$ and frequencies scale with $t^{-1}_{\mathrm{g}}$, when $M_{\mathrm{BH}}$ is increased the time-lag spectrum is uniformly stretched and squeezed in the horizontal and vertical direction, respectively. The dependence of the time-lags on $M_{\mathrm{BH}}$ is thus very similar to their dependence on $h$.

The bottom panel in Fig.\,\ref{figd1} shows the time-lag spectrum for various values of the outer disc radius. As $r_{\mathrm{out}}$ increases, $\tau_{\mathrm{plateau}}$ increases in magnitude and $\nu_{\mathrm{plateau}}$ decreases. The time-lag spectrum remains unaffected by $r_{\mathrm{out}}$ at frequencies higher than $\nu_{\mathrm{plateau}}$. In other words, $r_{\mathrm{out}}$ sets the level of the constant plateau at low frequencies, while this plateau occurs at increasingly lower frequencies as $r_{\mathrm{out}}$ increases. Our results are in agreement with CY15, who reported a similar dependence of the magnitude and shape of reverberation time-lag spectra on $r_{\mathrm{out}}$. The frequency $\nu_{\mathrm{plateau}}$ is proportional to $t^{-1}_{\mathrm{max}}$, and thus depends on $h$ and  $r_{\mathrm{out}}$.

Given our discussion above, it is clear that $\nu_{\mathrm{plateau}}$ and $\tau_{\mathrm{plateau}}$, which are the most pronounced features in the theoretical time-lag spectra, depend on $h$, $M_{\mathrm{BH}}$ and $r_{\mathrm{out}}$. Observationally, $\tau_{\rm plateau}$ appears to depend mainly on $M_{\rm BH}$ \citep[e.g.][]{2013MNRAS.431.2441D}, which implies that $h$ and $r_{\mathrm{out}}$ should be approximately the same in all AGN. Even so, the normalisation of this relation cannot directly indicate the height of the X-ray source, as $\tau_{\mathrm{plateau}}$ also depends on $r_{\mathrm{out}}$. The dependence is not as strong as that on $h$, but is present nevertheless. A more detailed theoretical study of the $M_{\mathrm{BH}}-\tau_{\mathrm{plateau}}$ relation is necessary before reaching conclusions based on the observed normalisation of this relation. This conclusion is strengthened by the fact that the discussion above is based on response functions estimated for the lamp-post geometry. If the X-ray source has a finite size, we expect the response function rise time to be altered. Since this directly affects $\tau_{\mathrm{plateau}}$, a study of the response functions from more complicated geometries is necessary to interpret the observations.

\section{Expected time-lag bias} \label{appe}

The best-fit model B parameters listed in Table\,\ref{table3} could significantly differ from their intrinsic values if the bias of the time-lag estimates has a magnitude comparable to, or larger than, their error. To investigate this possibility, we estimated the time-lag bias for $\{a,h\}=\{1,2.3r_{\mathrm{g}}\}$ and $\{0,100r_{\mathrm{g}}\}$ (we assumed $\theta=40^{\circ}$, $M_{\mathrm{BH}}=2\times10^6\,\mathrm{M}_{\odot}$ and $r_{\mathrm{out}}=10^3r_{\mathrm{g}}$ in both cases). These are two extreme cases in the parameter space we considered. 

\citet{2013MNRAS.435.1511A} were the first to quantify the effects of windowing on the time-lag bias. They showed that the time-lag bias can be up to $\sim30\%$ of the intrinsic value. EP16 also studied these effects in detail by exploring a wider parameter space. They showed that a model CS (not just a model time-lag spectrum) needs to be prescribed to estimate the time-lag bias, as the bias is introduced to the cross-periodogram itself. We assumed a model CS given by Eq.\,\ref{eqa3} and that there are no delays between variations of different energy bands in the continuum. To determine the amplitude of the model CS, we assumed that a) the continuum PSD in both energy bands is equal to the characteristic bending power-law shape observed in the X-ray light curves of many AGN \citep[e.g.][]{2012A&A...544A..80G}, and b) the intrinsic coherence between the two energy bands is unity at all frequencies. This uniquely determines the intrinsic continuum CS, $C^{(\mathrm{c})}_{5-7,2-4}(\nu)$, appearing in Eq.\,\ref{eqa3}, which is then given by
\noindent
\begin{equation} \label{eqe1}
C^{(\mathrm{c})}_{5-7,2-4}(\nu)=\frac{\mathscr{A}\nu^{-1}}{1+(\nu/\nu_{\mathrm{b}})^{\alpha-1}},
\end{equation}
\noindent
where $\mathscr{A}$ is the amplitude, $\alpha$ the high-frequency slope, and $\nu_{\mathrm{b}}$ the so-called bend-frequency. The typical values for these parameters are $\mathscr{A}\sim0.01$ (in so-called root-mean-square units), $2\lesssim\alpha\lesssim3$, and $\nu_{\mathrm{b}}\sim10^{-5}-10^{-4}\,\mathrm{Hz}$ for $M_{\mathrm{BH}}\sim10^6-10^7\,\mathrm{M}_{\odot}$. We therefore considered the cases $\{\alpha,\nu_{\mathrm{b}}\}=\{2,2\times10^{-4}\,\mathrm{Hz}\}$, $\{3,2\times10^{-4}\,\mathrm{Hz}\}$ and $\{2,2\times10^{-5}\,\mathrm{Hz}\}$ that are appropriate for our sample. We furthermore set $\mathscr{A}=0.01$, as the PSD amplitude was found by EP16 to not affect the time-lag bias.

\begin{figure*}[ht]
\centering
\includegraphics[width=15.0cm,clip]{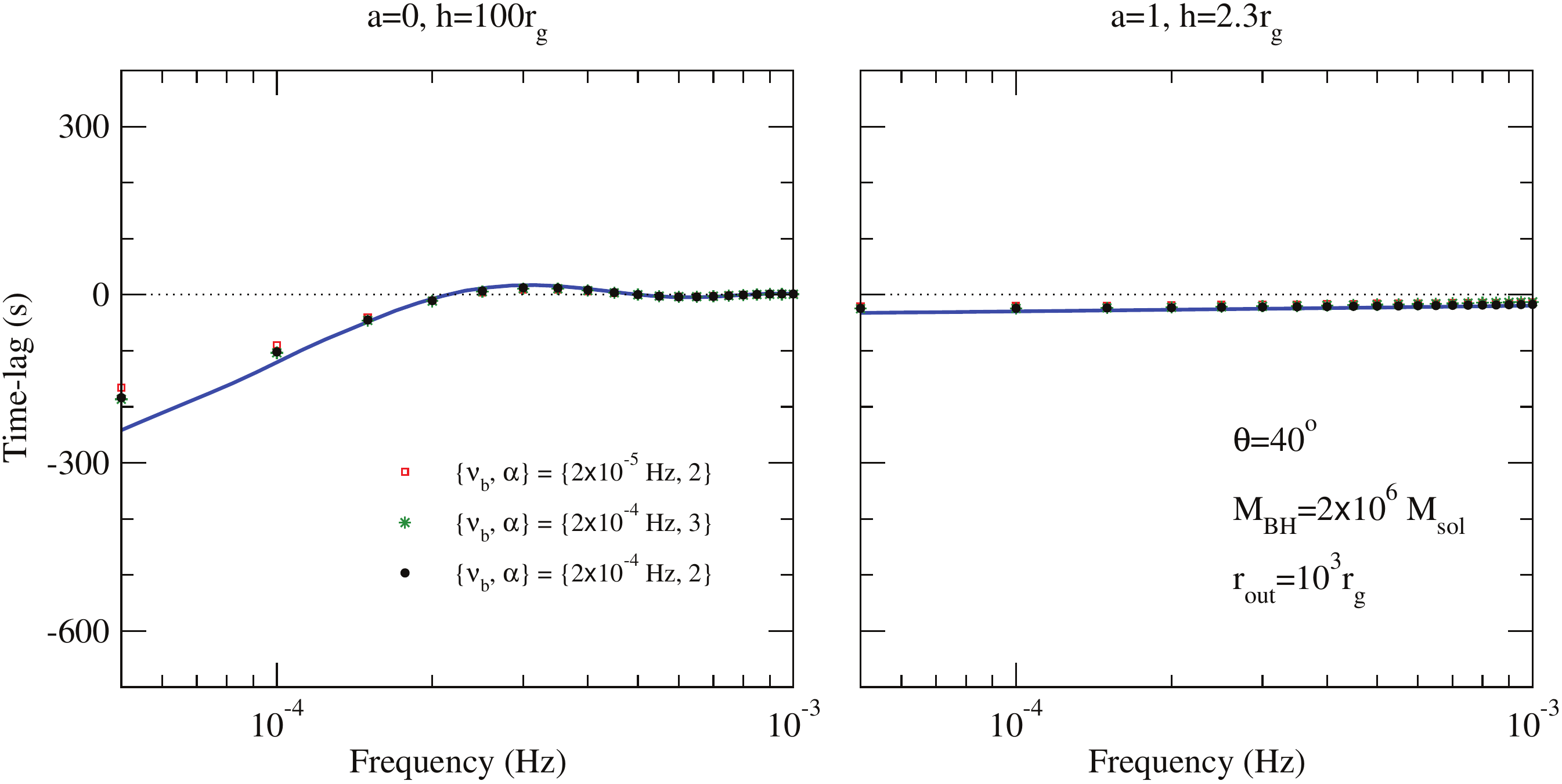}
\caption{Expected mean sample time-lag spectra computed from $20\,\mathrm{ks}$ segments for various model CS parameters (see the text for more details). The continuous blue curve indicates the model time-lag spectrum in each case.}
\label{fige1}
\end{figure*}

Given our model CS, we finally determined the expected mean of the time-lag estimates computed from $20\,\mathrm{ks}$ segments using Eq.\,\ref{eq13} in EP16. The results are shown in Fig.\,\ref{fige1}. The continuous blue line in the two panels indicates the model time-lag spectrum. Filled black circles, open red squares, and green stars correspond to the expected mean sample time-lag spectrum for different $\{\alpha,\nu_{\mathrm{b}}\}$ values (as noted in the figure).

The horizontal axis indicates the widest frequency range for which we were able to obtain reliable time-lag estimates using real data. The range of values in the vertical axis is identical to the corresponding range in Figs.\,\ref{fig2} and \ref{fig3} (with the exception of Ark 564). The difference between each point and the corresponding model at a given frequency in Fig.\,\ref{fige1} is equal to the expected time-lag bias, whose magnitude needs to be compared to the error bars in Figs.\,\ref{fig2} and \ref{fig3}. As noted by EP16, the mean of the time-lag estimates is always smaller (in magnitude) than their intrinsic values at each frequency.

For low X-ray source height values, it is clear that the magnitude of the bias is entirely negligible compared to the time-lag errors and hence should not affect our results. For high X-ray source height values the bias is more significant at low ($\lesssim10^{-4}\,\mathrm{Hz}$) frequencies, although still smaller than the time-lag errors for all the sources in our sample. Perhaps in this case the best-fit height values may slightly underestimate their intrinsic values, although this effect should not be significant.

\end{appendix}

\end{document}